%% file: main.tex
\documentclass[aps, prd, twocolumn, superscriptaddress, longbibliography,nofootinbib]{revtex4-1}
\pdfoutput=1

\input{header}

\begin{document}

\title{Revisiting Primordial Black Hole Capture by Neutron Stars}

\author{Roberto Caiozzo\!\orcidB{}}
\email{rcaiozzo@sissa.it}
\FirstAffiliation
\SecondAffiliation

\author{Gianfranco Bertone\!\orcidA{}}
\SecondAffiliation

\author{Florian K{\"u}hnel\!\orcidC{}}
\ThirdAffiliation
\FourthAffiliation

\date{\formatdate{\day}{\month}{\year}, \currenttime}

\begin{abstract}
\noindent
\input{abstract}
\end{abstract}

\maketitle

\section{Introduction}
\label{sec:Introdcution}
\input{intro}
\section{Path of PBHs and Energy Loss}
\label{sec:Section A}
\input{Energy_loss_formalism}

\section{Energy Loss for a realistic Neutron Star Interior}
\label{sec:Section-B}
\input{Realistic}

\section{Central Parsec}
\label{sec:Section-C}
\input{Central_parsec}

\section{Capture rate}
\label{sec:Section-D}
\input{Capture}

\section{Time to Collapse}
\label{sec:Section-E}
\input{Time}

\section{Expected Collapse Number and Survival Likelihoods}
\label{sec:Section-F}
\input{Collapses}

\section{Constraints and Dense Environment Effects}
\label{sec:Section-G}
\captionsetup{width=.95\textwidth}
\begin{figure*}[ht]
\includegraphics[width=0.92\textwidth]{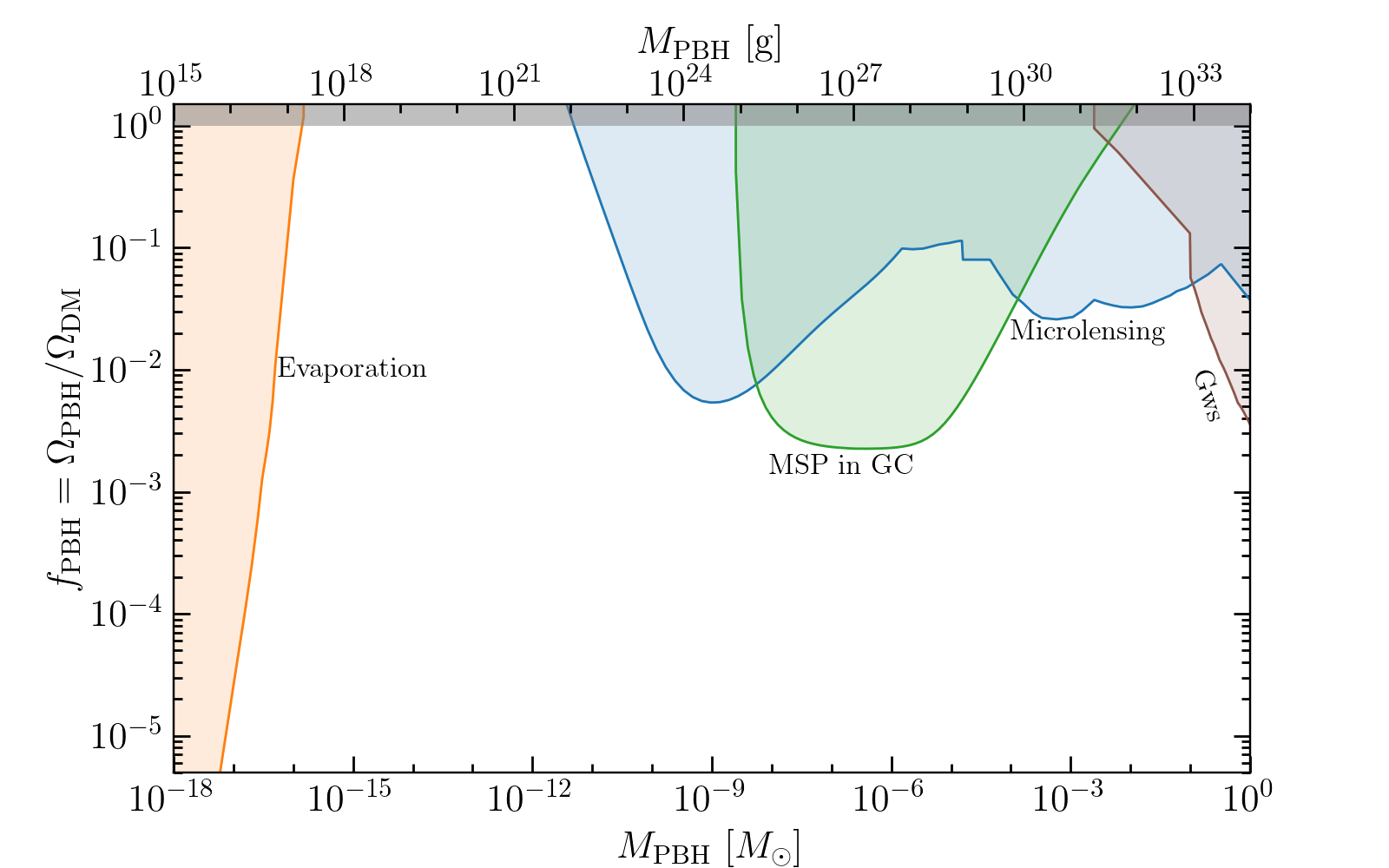}
    \caption{Constraints on the PBH abundance arising from the most optimistic case of an observation of a single $2.12\,M_{\odot}$ MSP at $10^{-3}\,$pc in the Galactic centre, assuming an adiabatically compressed dark matter profile with a spike around Sgr A* (green). Existing constraints on the PBHs abundance are also shown for reference. Data and code for this figure are taken from Ref.~\cite{Green:2020jor}.}
    \label{opt}
    \end{figure*}

\input{Constraints}

\begin{figure*}[p]
    \includegraphics[width=0.45\textwidth]{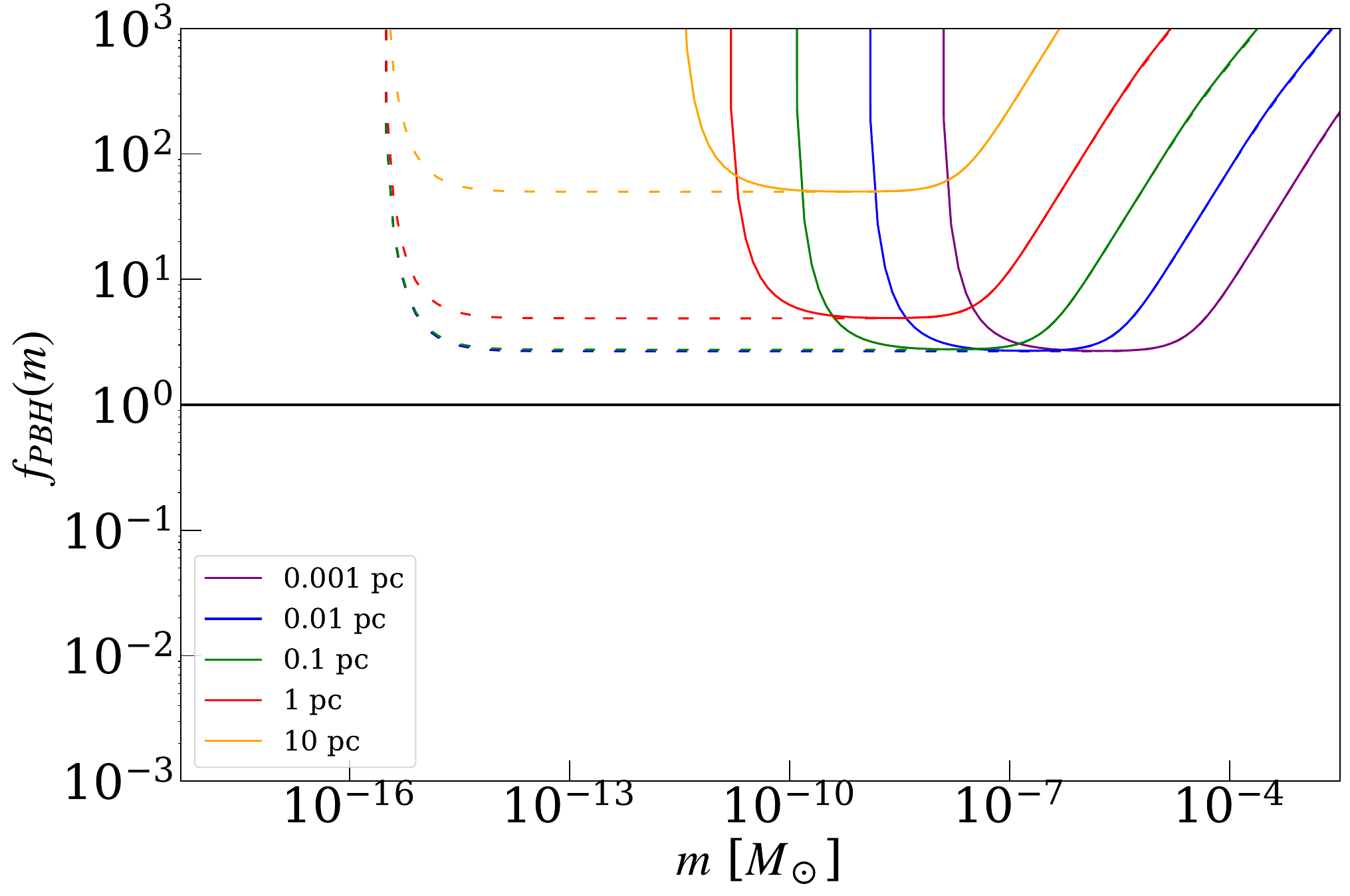}
    \includegraphics[width=0.45\textwidth]{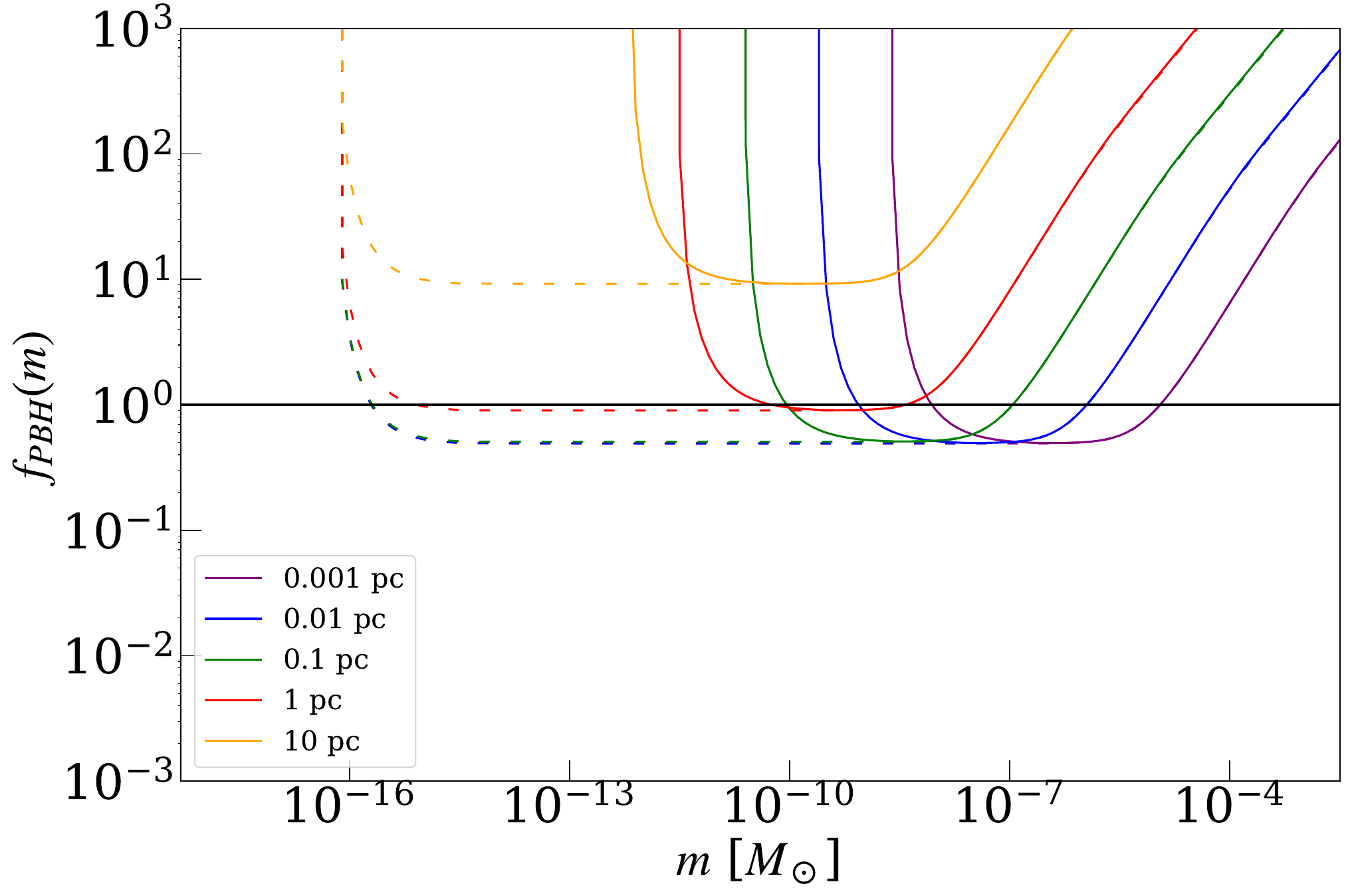}
    \\
    \includegraphics[width=0.45\textwidth]{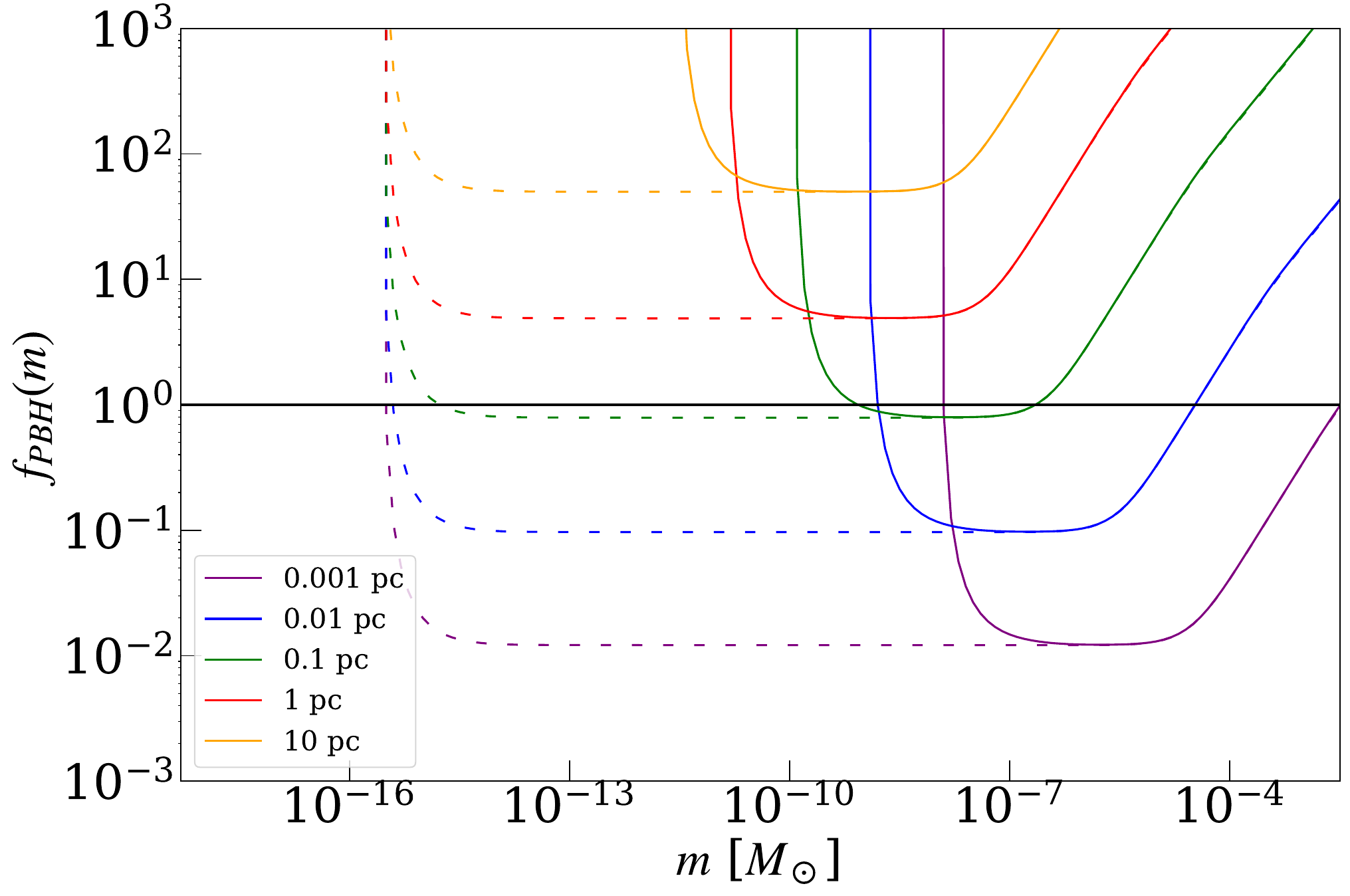}
    \includegraphics[width=0.45\textwidth]{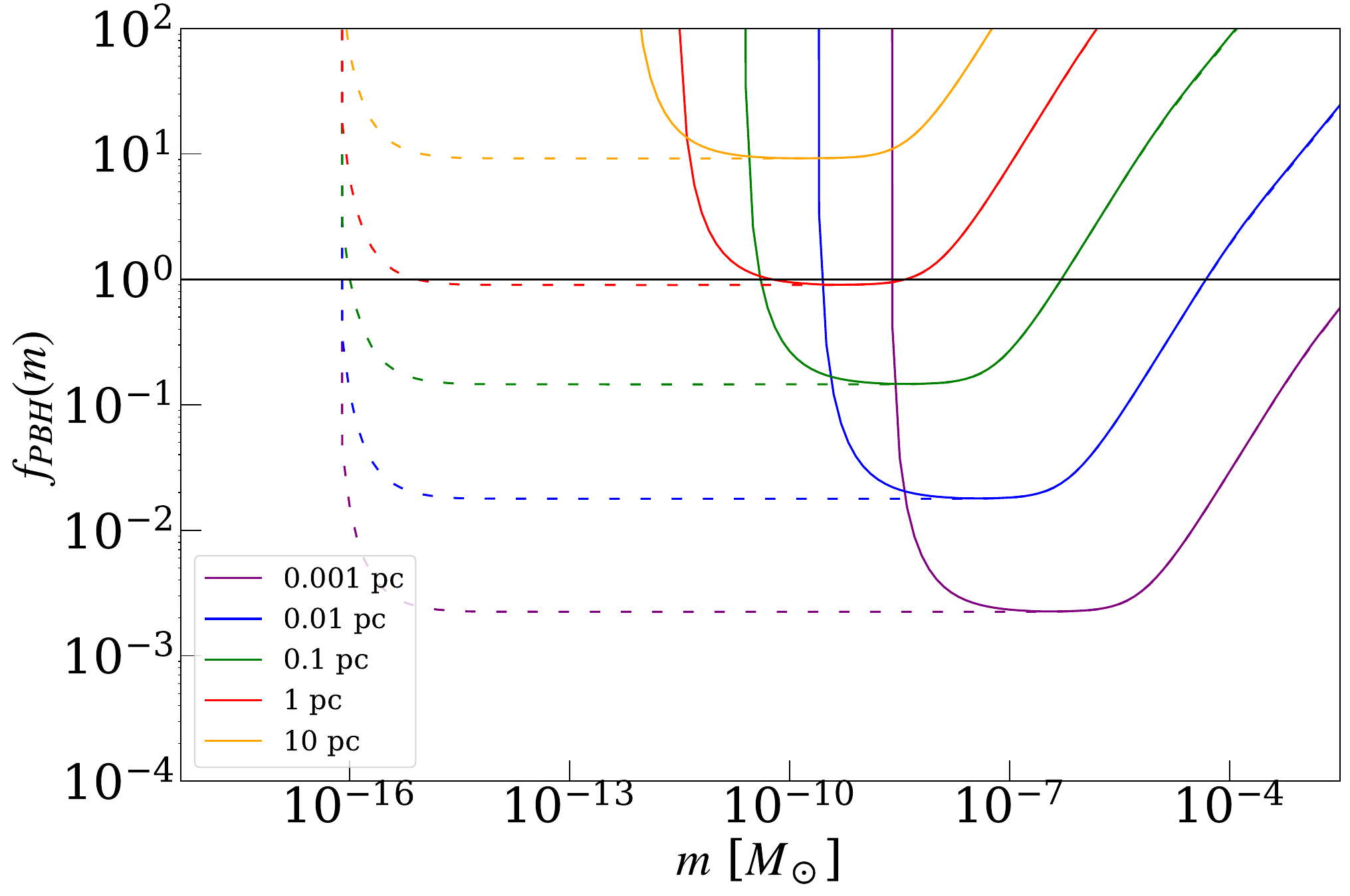}
    \\
    \includegraphics[width=0.45\textwidth]{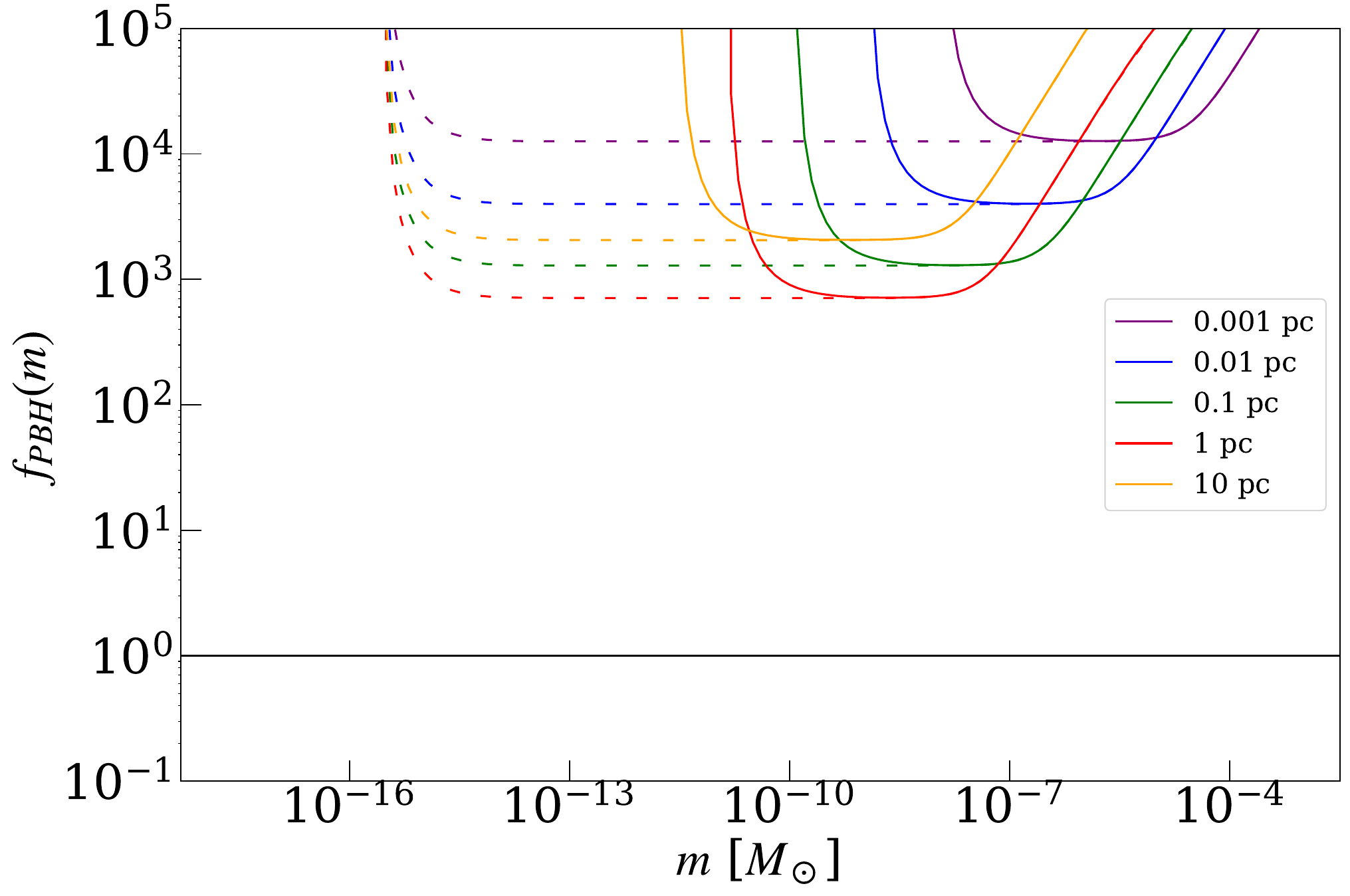}
    \includegraphics[width=0.45\textwidth]{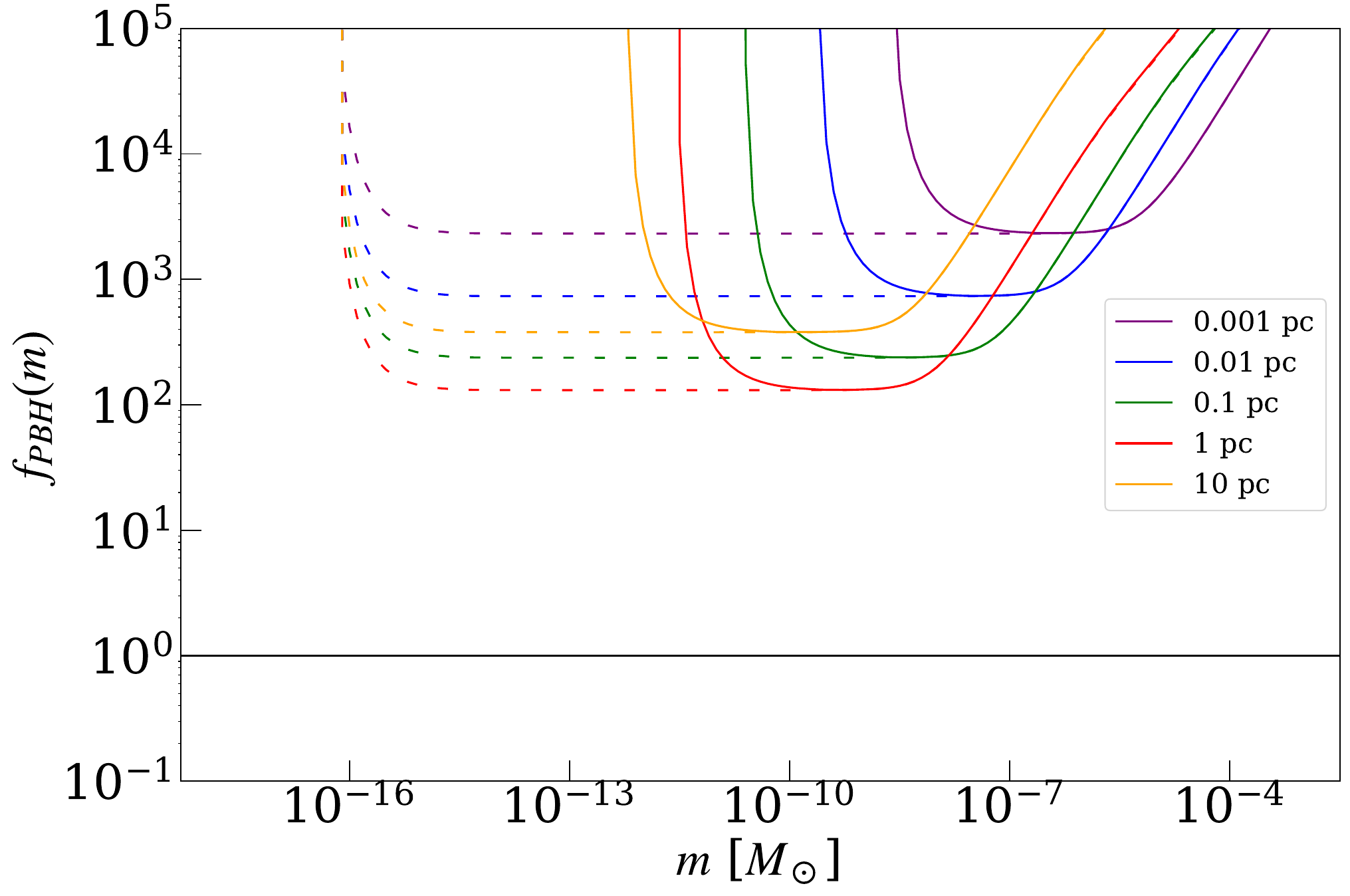}
    \\
    \includegraphics[width=0.45\textwidth]{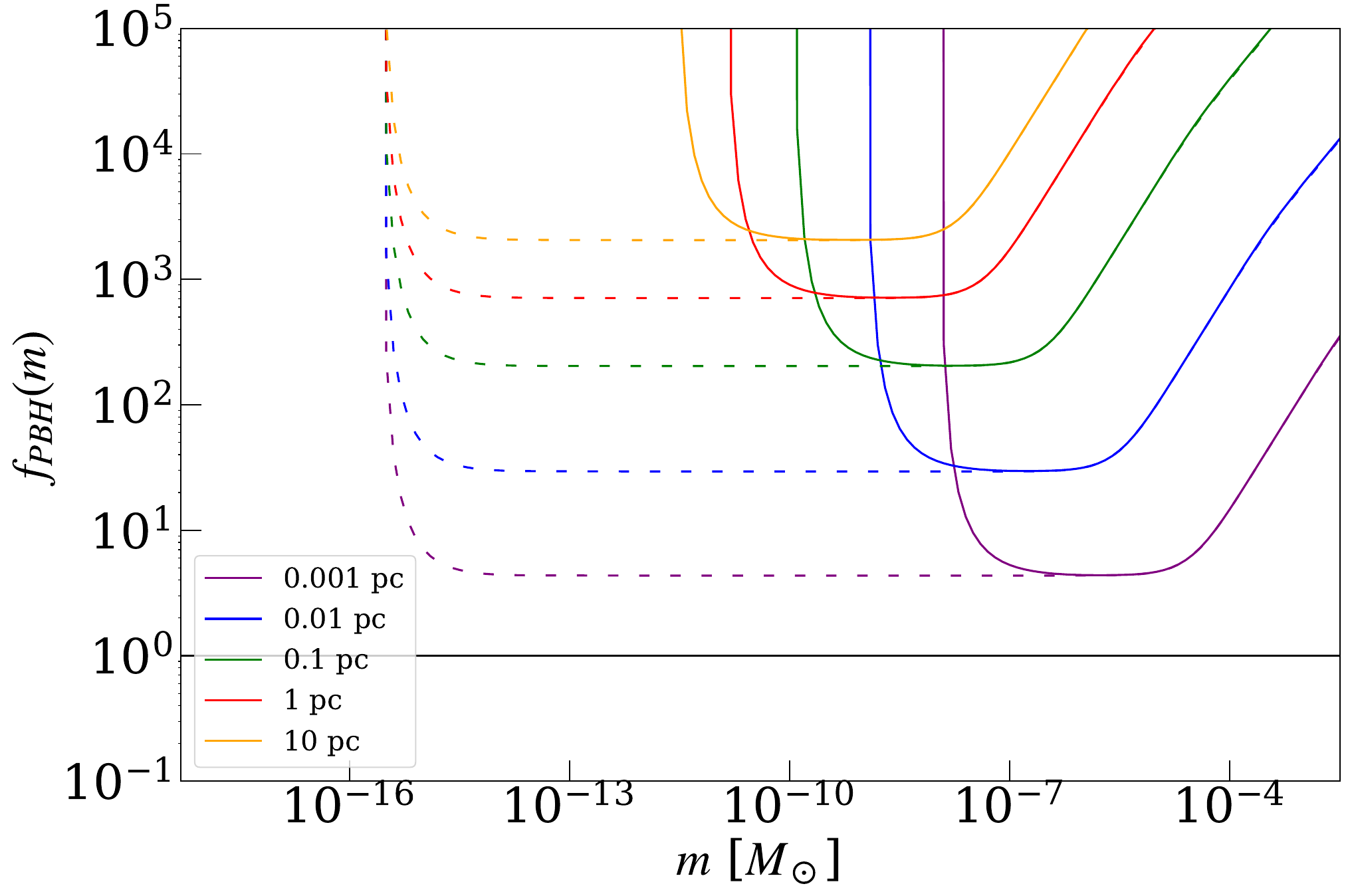}
    \includegraphics[width=0.45\textwidth]{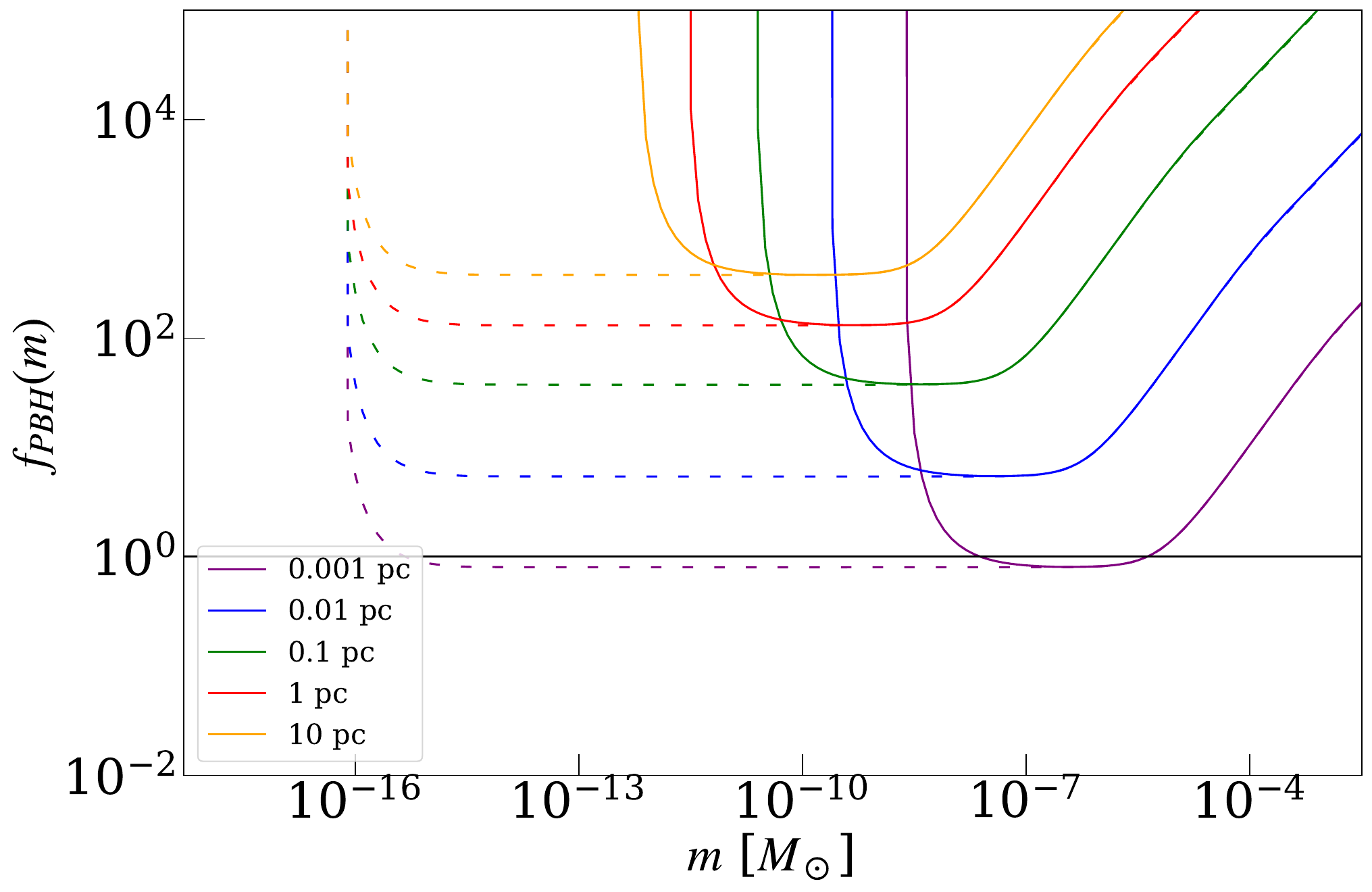}
    \caption{Constraints from observing a $1.52\,M_{\odot}$ (left) or a $2.12\,M_{\odot}$ (right) MSP at different distances from the Galactic centre (as shown in the legends) for 
        an adiabatic dark matter profile (first row), 
        a spiked Adiabatic profile (second row), 
        an NFW profile (third row), and
        a spiked NFW profile (fourth row).
    Dotted lines indicate constraints ignoring dense environment effects on orbital decay.}
    \label{all}
    \end{figure*}

\section{Conclusions}
\label{sec:Discussion-and-Outlook}
\input{conclusion}

\textit{Acknowledgements.} GB gratefully acknowledges the Department of Physics of Columbia University and the Italian Academy for Advanced Studies in America, where part of this work was carried out. RC also thanks Dr.~Oscar Macias Ramirez for his help in sorting through the pulsar literature.
\newpage

\appendix
\section{Survival likelihood for a population of NSs}
\label{A}
\input{appendix}

\section{A useful transformation}
\label{B}
\input{appendix2}

\setlength{\bibsep}{4pt}
\bibliography{References}

\end{document}

%% file: header.tex
\usepackage{amsmath}	
\usepackage{amsthm}		
\usepackage{amssymb}	

\usepackage{datetime}
\usepackage{graphicx}
\usepackage{color}
\usepackage{verbatim}
\usepackage{bm}
\usepackage{braket}
\usepackage{cancel}
\usepackage{xcolor}
\usepackage{slashed}
\usepackage{braket}
\usepackage{caption}
\usepackage{subcaption}
\usepackage{amsfonts} 
\usepackage{float}
\usepackage{dsfont}
\smallskipamount
\usepackage{feynmp}
\usepackage{letltxmacro} 
\usepackage{microtype}

\usepackage[colorlinks=true, citecolor=midblue, linkcolor=midblue, urlcolor=midblue]{hyperref}

\usepackage{setspace}

\usepackage{tikz,xcolor}
\definecolor{lime}{HTML}{A6CE39}
\DeclareRobustCommand{\orcidicon}{
	\begin{tikzpicture}
	\draw[lime, fill=lime] (0,0) 
	circle [radius=0.16] 
	node[white] {{\fontfamily{qag}\selectfont \tiny ID}};
	\draw[white, fill=white] (-0.0625,0.095) 
	circle [radius=0.007];
	\end{tikzpicture}
	\hspace{-2mm}
}
\foreach \x in {A, ..., Z}{
	\expandafter\xdef\csname orcid\x\endcsname{\noexpand\href{https://orcid.org/\csname orcidauthor\x\endcsname}{\noexpand\orcidicon}}
}

\settimeformat{ampmtime}

\definecolor{grey}{rgb}{0.4,0.4,0.4}
\definecolor{dullmagenta}{rgb}{0.4,0,0.4}
\definecolor{darkblue}{rgb}{0,0,0.4}
\definecolor{midblue}{rgb}{0,0,0.5}
\definecolor{midred}{rgb}{0.5,0,0}
\definecolor{orange}{rgb}{1,0.5,0}
\definecolor{lightbrown}{rgb}{0.75,0.5,0.25}
\definecolor{tan}{cmyk}{0.14,0.42,0.56,0}
\definecolor{djunglegreen}{cmyk}{0.99,0,0.52,0}
\definecolor{lightgreen}{rgb}{0,1,0}
\definecolor{olivegreen}{cmyk}{0.64,0,0.95,0.40}
\definecolor{midgreen}{rgb}{0.0,0.675,0.0}
\definecolor{darkgreen}{rgb}{0,0.5,0}

\newcommand{\vs}{\vspace}

\renewcommand{\.}{\hspace{0.5mm}}

\newcommand{\Crm}{\ensuremath{\mathrm{C}}}

\newcommand{\crm}{\ensuremath{\mathrm{c}}}
\newcommand{\drm}{\ensuremath{\mathrm{d}}}

\newcommand{\prm}{\ensuremath{\mathrm{p}}}

\newcommand{\rrm}{\ensuremath{\mathrm{r}}}
\newcommand{\srm}{\ensuremath{\mathrm{s}}}

\newcommand{\Ccal}{\ensuremath{\mathcal{C}}}

\newcommand{\Gcal}{\ensuremath{\mathcal{G}}}

\renewcommand{\d}{\ensuremath{\mathrm{d}}}

\newcommand{\cf}{c.f.}

\settimeformat{ampmtime}

\usepackage{float}

\makeatletter
\let\oldr@@t\r@@t
\def\r@@t#1#2{%
\setbox0=\hbox{$\oldr@@t#1{#2\,}$}\dimen0=\ht0
\advance\dimen0-0.2\ht0
\setbox2=\hbox{\vrule height\ht0 depth -\dimen0}%
{\box0\lower0.4pt\box2}}
\LetLtxMacro{\oldsqrt}{\sqrt}
\renewcommand*{\sqrt}[2][\ ]{\oldsqrt[#1]{#2}}
\makeatother
\usepackage[normalem]{ulem}

\usepackage{caption} 

\usepackage{soul}

\newcommand{\FirstAffiliation}{\affiliation{
International School for Advanced Studies (SISSA), 
     Via Bonomea, 265, 34136 Trieste TS, Italy
}}
 
\newcommand{\SecondAffiliation}{\affiliation{Gravitation Astroparticle Physics Amsterdam (GRAPPA), 
    University of Amsterdam, 
    Science Park 904, 1098 XH Amsterdam, 
    The Netherlands
	}}

\newcommand{\ThirdAffiliation}{\affiliation{
	Max-Planck-Institut f{\"u}r Physik,
	Boltzmannstr.~8,
    85748 Garching,
	Germany}}

\newcommand{\FourthAffiliation}{\affiliation{
	Arnold Sommerfeld Center,
	Ludwig-Maximilians-Universit{\"a}t,
	Theresienstra{\ss}e 37,
	80333 M{\"u}nchen,
	Germany}}
 

%% file: abstract.tex
A sub-solar mass primordial black hole (PBH) passing through a neutron star, can lose enough energy through interactions with the dense stellar medium to become gravitationally bound to the star. Once captured, the PBH would sink to the core of the neutron star, and completely consume it from the inside. In this paper, we improve previous energy-loss calculations by considering a realistic solution for the neutron star interior, and refine the treatment of the interaction dynamics and collapse likelihood. We then consider the effect of a sub-solar PBH population on neutron stars near the Galactic center. We find that it is not possible to explain the lack of observed pulsars near the galactic center through dynamical capture of PBHs, as the velocity dispersion is too high. We then show that future observations of old neutron stars close to Sgr A* could set stringent constraints on the PBHs abundance. These cannot however be extended in the currently unconstrained asteroid-mass range, since PBHs of smaller mass would lose less energy in their interaction with the neutron star and end up in orbits that are too loosely bound and likely to be disrupted by other stars in the Galactic center.

%% file: intro.tex
Primordial black holes (PBHs) have long been recognized as viable dark matter candidates~\cite{Hawking:1971ei, Zeldovich:1967lct, Carr:1974nx} (for recent reviews, see e.g.~Refs.~\cite{Carr:2016drx, Carr:2020xqk, Green:2020jor, Escriva:2022duf}). The interest in PBHs goes beyond their cosmological relevance, as their presence has been invoked to explain a variety of astrophysical observations (\cf~Ref.~\cite{Carr:2023tpt}), such as 
{\it 1)} microlensing events by planetary-mass objects 
~\cite{Niikura:2019kqi}; 
{\it 2)} microlensing of quasars
~\cite{Mediavilla:2017bok}; 
{\it 3)} microlensing events by objects between 2 and 5 solar masses~\cite{Wyrzykowski:2019jyg}; 
{\it 4)} correlations in the X-ray and cosmic infrared background fluctuations~\cite{Kashlinsky:2005di};
{\it 5)} non-observations of certain ultrafaint dwarf galaxies~\cite{Clesse:2017bsw}; 
{\it 6)} masses, spins and coalescence rates for black holes found by LIGO/Virgo~\cite{LIGOScientific:2018mvr}; 
{\it 7)} relationship between the mass of a galaxy and that of its central supermassive black hole~\cite{Kruijssen:2013cna}; 
{\it 8)} so-called $G$ objects near the Galactic centre (GC)~\cite{Flores:2023lll, Takhistov_2021}. Many constraints exist on the PBH abundance~\cite{Carr:2016drx, Carr:2020xqk, Green:2020jor, Escriva:2022duf}. Viable scenarios however exist in which PBHs can in principle be compatible with all constraints and observations~\cite{Carr:2020xqk}, for instance when the change in degrees of freedom as the Universe cools from the QCD scale to the electron mass leads to a multi-modal PBH mass function~\cite{Carr:2019kxo}.

In this article, we assess constraints on the abundance of sub-stellar mass PBHs, by performing a state-of-the-art analysis of dynamical PBH capture by neutron stars (NS), extending and refining previous studies~\cite{Capela_Pshirkov_Tinyakov_2013, Takhistov:2017bpt, Génolini_Serpico_Tinyakov_2020, Abramowicz_2018, Abramowicz_2022, Fuller:2017uyd}. Once captured, the PBH sinks into the core of the NS, accretes the stellar matter, and leaves behind a black hole of mass around that of the initial NS. 

We specifically focus on the innermost parsec of our Milky Way. This region, which we refer to as the Galactic center (GC), is of particular interest for the process of PBH capture by NSs, due to both the expected high density of dark matter and the predicted large population of pulsars~\cite{Pfahl_2004, Gondolo_1999}. It is estimated that thousands of radio pulsars should be present in the GC~\cite{Pfahl_2004}. However, a number of surveys have struggled to find any within the innermost $25\,$parsec~\cite{Dexter_2014}. This was thought to be mainly due to strong temporal scatterings, altering the discrimination of pulsar signals from the inner parsecs~\cite{Cordes_1997}. However, the 2013 discovery of the magnetar SGR J1745-29 within $0.1\,$pc from Sagittarius A*~\cite{Mori_2013}, suggests that the temporal scattering cannot be as strong as previously thought~\cite{Spitler_2013}. Several mechanisms have been put forward to explain this missing-pulsar problem in the GC, including efficient magnetar formation, and disruption of the expected pulsar population~\cite{Dexter_2014}. We investigate here the possibility of NS disruption by PBHs (see also Ref.~\cite{Fuller:2017uyd}).

Millisecond pulsars (MSPs) would be excellent candidates for PBH capture at the GC. Their presence could hence be used to formulate abundance constraints on PBHs. Because of their energy spectrum, MSPs are a likely candidate for the observed GeV excess from the Galactic bulge~\cite{Calore_Mauro_Donato_Hessels_Weniger_2016}, which suggests a large population of these objects extending into the Galactic center. Ref.~\cite{Generozov_Stone_Metzger_Ostriker_2018} predicts from numerical simulations 67 long lived NS-Xray binaries in the innermost parsec of the Milky Way. Given that these binaries are the progenitor of millisecond pulsars, it is reasonable to expect a population of MSPs extending in the central parsec. Despite the fact that their detection in GC is technically challenging, the temporal scattering measured from the SGR J1745-29 suggests that such a detection might be possible~\cite{Dexter_2014}. 

In the article, we investigate the rate of dynamical capture of PBHs by NSs in the GC, revising and extending existing calculations, considering in particular more realistic star interiors. We calculate the likelihood of collapse for a NS of a given age in that region, and assess the disruption rate of pulsars in the GC. 

The paper is structured as follows: in Section~\ref{sec:Section A} we discuss the formalism for energy loss of a PBH crossing a NS. In Section~\ref{sec:Section-B} we explain the NS interior solution used for our calculations and the effect this has on the energy loss. Section~\ref{sec:Section-C} explores the relevant properties of the GC. In Section~\ref{sec:Section-D} we explain how to calculate the PBH capture rate from NSs. We then expand on how to calculate the likelihood of collapse for a NS, accounting for the time of orbital collapse for the PBH (Section~\ref{sec:Section-E}) as well as dense environment effects (Section~\ref{sec:Section-F}), and how to use this to obtain constraints on the abundance of PBHs from the observation of old NS in the GC (Section~\ref{sec:Section-G}). Throughout this paper we use units in which $c = 1$.

%% file: Energy_loss_formalism.tex
\begin{table}[b]
    \centering
    \begin{tabular}{c|l|l}
    \hline \hline 
    $i$&\text {$\phi_i$} & \text {$\psi_i$}  \\
    \hline 
    0\; &\;$\cos^{-1}(-\frac{1}{e})$&\;$\psi_{0}-\cos^{-1}\!\left(\frac{\Tilde{b}\sqrt{e^{2}-1}-1}{e}\right)$\\
    1\; &\;$\phi_{0}-\cos^{-1}\!\left(\frac{\sqrt{1-\Tilde{\alpha}_-^{2}}}{\varepsilon}\right)$&\;$\pi+2\psi_{1}-\phi_{0}$\\
    2\;&\;$\pi+2\psi_{1}-\psi_{0} $&\;$\pi+2\psi_{1}$ \\
    \hline \hline
    \end{tabular}
    \caption{Angles for the orbital formalism.}
    \label{tab:angles}
\end{table}
In order to calculate the energy lost by the PBH through gravitational interaction with the NS, we start by deriving an expression for the PBH trajectory inside and outside the star. Following previous work from G{\'e}nolini {\it et al.}~\cite{Génolini_Serpico_Tinyakov_2020}, we use the Newtonian approximation and assume that the path of the PBH is not significantly perturbed by the NS interactions during the first crossing. This is a valid assumption since the kinetic energy of the PBH in the star is much larger then the energy lost by the PBH during one crossing.

{\bf Path of PBHs.} For a NS of mass $M$ and radius $R$, all possible approach orbits are fully characterized by just two parameters: the asymptotic velocity at infinity, $v_i$, and the impact parameter $b$. The path of the PBH outside the star is described by a hyperbolic orbit, which is parameterized by a semi-minor axis equal to the impact parameter and a semi-major axis $a$ and eccentricity $e$ given by 
\begin{equation}
    a
        =
            \frac{GM}{v_i^{2}}
    \, ,
    \quad
    e
        =
            \sqrt{
                1+\frac{\Tilde{b}^{2} v_i^{4}}{v_{\star}^{4}}
            }
            , 
\end{equation}
where $\Tilde{b} = b/R$ is the reduced impact parameter and $v_{\star} = (GM / R)^{1/2}$ is the typical velocity of the star. 

If $b$ is smaller then the critical impact parameter $b_{\crm}=R (1+2 v_{\star}^{2} / v_i^{2})^{1/2}$, the PBH will cross the NS and a new expression becomes necessary to describe the path of the PBH inside the star. Approximating the NS as a sphere of constant density, the path of the PBH, while inside the star, will be described by an ellipse centered at the NS core, being parameterized by a semi-major and semi-minor axis given by
\begin{equation}
    \alpha_{\pm}
        =
            R\,\sqrt{\frac{3}{2}\left(1 \pm \sqrt{1-\left(\frac{2v_i \tilde{b}}{v_{\star} 3}\right)^{2}}\right)}
            \; .
\end{equation}
Patching the two regimes together, we get~\cite{Génolini_Serpico_Tinyakov_2020}
 \begin{equation}
    r(\phi)
        =
            \begin{cases}
                \frac{a\left(e^{2}-1\right)}{1+e \cos\!\left(\phi-\psi_{0}\right)},
                    & 0< \phi \leqslant \phi_{0}
                    \\[2mm]
                \frac{\alpha_{-}}{\sqrt{1-\varepsilon^{2} \cos ^{2}\left(\phi-\psi_{1}\right)}}\, ,
                    & \phi_{0}<\phi<\phi_{1}
                    \\[2mm]
                \frac{a\left(e^{2}-1\right)}{1+e \cos\!\left(\phi-\psi_{2}\right)}\, ,
                        & \phi_{1} \leqslant \phi<\phi_{2}
                        \, ,
            \end{cases}
\end{equation}
where the angles $\phi_i$ and $\psi_i$ are given in Table~\ref{tab:angles} and $\varepsilon = \sqrt{1- \alpha_{-} / \alpha_{+}}$ is the eccentricity of the ellipse parametrizing the internal path.

This is the same parametrization as used in Ref.~\cite{Génolini_Serpico_Tinyakov_2020}, but with corrected typographical errors in the equations for the total deflection angle, the semi-major and semi-minor axis inside the star.

{\bf Energy losses.} Black holes, if not electrically charged, can only interact gravitationally with surrounding matter, and PBHs are no exception. Assuming that charge and angular momentum are negligible, as recent analyses~\cite{Araya_2023, Luca_2019} seem to suggest for a large class of PBH formation scenarios\footnote{Note that in some scenarios PBH do have a large spin, e.g. if they form during matter-domination~\cite{Khlopov:1985fch}, from Q-balls~\cite{Cotner:2016cvr}, scalar-field fragmentation~\cite{Cotner:2019ykd}, oscillons~\cite{Cotner:2018vug}, or quark confinement~\cite{Dvali:2021byy}.}, there are only four relevant mechanisms for a PBH to lose energy while orbiting close or through a NS:
\begin{itemize}
    \itemsep0em
    
    \item gravitational waves (GW) emission;
    
    \item dynamical friction (DF);
    
    \item accretion of NS matter onto the PBH; and 
    
    \item production of surface waves (SW). 
\end{itemize}
If the total energy lost, $\Delta E_{tot}$, through the initial hyperbolic orbit is larger than the initial energy $E_i=\frac{1}{2}mv_{i}^{2}$, the PBH will be stuck in a bound orbit and, excluding external forces, will eventually sink to the NS's center and consume it~\cite{Capela_2013, Capela_2013-2, Kouvaris_2011}.
\newpage

{\bf Gravitational Waves emission.} A PBH orbiting close or passing through a NS, will lose energy due to GW emission. Given a PBH of mass $m$, the GW energy loss per unit angle outside the NS is given by~\cite{De_Vittori_2012}
\begin{equation}
\begin{aligned}
    p_{\phi}^{\rm out}
        &=
            \frac{m^{2}v_{\star}^{2}v_i^{5}}{45M}\frac{2}{\tilde{b}} \frac{(1+e \cos \phi)^{2}}{\left(e^{2}-1\right)^{3}}
            \times
            \\[2mm]
        &\hphantom{=\;\;}
            \big(
                144+288 e \cos \phi + 77 e^{2}+67 e^{2} \cos 2 \phi
            \big)
            \, ,
\end{aligned}
\end{equation}
and inside the NS it will be~\cite{Génolini_Serpico_Tinyakov_2020}
\begin{equation}
\begin{aligned}
    p_{\phi}^{\rm in}
        =
            \frac{m^{2}v_{\star}^{2}v_i^{5}}{45M}4 \frac{\tilde{b}^{5}}{\tilde{\alpha}_{-}^{6}}  \frac{\left(1-\varepsilon^{2}\right)^{2}}{\left(1-\varepsilon^{2} \cos ^{2} \phi\right)^{3}}\times\\[2mm]
            \left(72\left(1-\varepsilon^{2}\right)+37 \varepsilon^{4}+36 \varepsilon^{2}\left(\varepsilon^{2}-2\right) \cos 2 \phi-\varepsilon^{4} \cos 4 \phi\right) 
            .
\end{aligned}
\end{equation}

The energy loss per unit angle for the whole orbit can then be expressed as a piece-wise function in the form
\begin{equation}
    p_{\phi}
        =
            \begin{cases}
                p_{\phi}^{\rm out }\left(\phi-\psi_{0}\right)
                ,
                    & 0 \leqslant \phi \leqslant \phi_{0}\\[2mm]
                p_{\phi}^{\rm in}\left(\phi-\psi_{1}\right)
                ,
                    & \phi_{0}<\phi<\phi_{1}\\[2mm]
                p_{\phi}^{\rm out}
                \left(\phi-\psi_{2}\right),
                    & \phi_{1} \leqslant \phi \leqslant \phi_{2}
                    \, ,
                \end{cases}
\end{equation}
It is then straightforward to compute the energy loss from gravitational waves for a given orbit as 
\begin{equation}
    \Delta E_{\mathrm{gw}}
        =
            \int_{0}^{\phi_{2}} p_{\phi} \; \mathrm{d}\phi
            \, .
\end{equation}
In the case where the PBH does not cross the star, this can be solved analytically as~\cite{De_Vittori_2012}
\begin{equation}
    \Delta E_{\mathrm{gw}}
        =
            \frac{8}{15}
            \frac{m^{2}}{M}\!
            \left(
                \frac{v_{\star}^{2}}{\tilde{b}v_i}
            \right)^{\!7}
            p(e)
            \, ,
    \label{GW_loss}
\end{equation}
with
\begin{align}
    p(e)
        =&
            \phantom{\Bigg(}
                \arccos\!
                \left[
                    -\frac{1}{e}
                \right]\!
                \left(
                    24 + 73\.e^{2}
                    +
                    \frac{37}{4}\.e^{4}
                \right)
            \\[2mm]
            &
            \phantom{\Bigg(}
                +\frac{\sqrt{e^{2}-1}}{12}\
                \big(
                    602 + 673\.e^{2}
                \big)
            \phantom{\Bigg)}
            \. .
\end{align}

{\bf Surface waves.} Tidal effects are mostly negligible in PBH-NS interactions where the PBH does not cross into the star~\cite{tidalforces}. However, if the PBH passes through the NS, ripples are induced on the surface, leading to an associated energy loss from the PBH. These ripples are referred to in the literature as surface waves, there is no exact solution for the associated energy loss from a realistic star, but Ref.~\cite{Defillon_2014} has shown that the result should be of order
\begin{equation}  
    \Delta E_{\rm SW}
        \approx
            3\frac{G m^{2}}{R}
            \, ,
\end{equation}
being a suitable approximation given that SW will be subdominant to accretion and dynamical friction for most of the impact parameters smaller than $b_{\crm}$.

{\bf Accretion.} While the PBH is inside the NS, some of the star's mass will be accreted by the black hole. This mass will have to be accelerated to the local PBH velocity, which will cause an accretion force in the direction opposite to the PBH's motion. The internal temperature of the NS at formation is around $10^{11}\,$K. The star then rapidly cools down through neutrino emissions, and during the lifetime of a pulsar, the internal temperature is $T\,( 10\,{\rm Myr} ) \sim 10^{7}\,$K~\cite{Yakovlev_2010}, with an associated thermal velocity $v_{\rm th} \approx 10^{-3}$. Given that the PBH velocity within the NS is much higher than the local thermal velocity of the neutrons, we treat them as static in the NS reference frame. In the limit $v\ll c_{\rm s}$ (PBH velocity much smaller then the sound speed), including relativistic effects, the rate of mass capture for a PBH with speed $v$ (with associated Lorentz factor $\gamma$) in a perfect fluid at equilibrium with density $\rho$ and pressure $P$ is~\cite{1989ApJ...336..313P}
\begin{equation}
    \dot{m}
        =
            \pi\.d_{\crm}^{2}\.
            ( \rho + P )\.\gamma\.v
            \, ,
\end{equation}
where $d_{\crm}$ is the critical impact parameter for a Schwarzschild black hole at the local PBH velocity, which is given by solving the set of equations~\cite{Capela_2013}
\begin{equation}
\begin{aligned}
    \gamma^{2}
        &=
            U(x)
            \, ,
            \\[1.5mm]
    \frac{\partial U}{\partial x}
        &=
            0\, ,
\end{aligned}
\end{equation}
where $U(x) = ( 1 + [ d\.v \gamma\.x / R_{\rm s} ]^{2} )( 1 - x )$ and $R_{\rm s}=2Gm$ is the Schwarzschild radius of the PBH. The collision with the captured mass can then be treated as perfectly inelastic, the consequent drag force experienced by the PBH will be 
\begin{equation}
    \bm{F}_{\mathrm{acc}}
        =
            -\pi d_{\crm}^{2}v^{2}\gamma^{2}(\rho + P)\frac{\bm{v}}{v}
            \, .
\end{equation}

Approximating the path of the PBH through the NS as described before, the total loss of kinetic energy is
\begin{equation}
    \Delta E_{\mathrm{acc}}
        =
            \int_{\phi_{0}}^{\phi_{1}}|F_{\mathrm{acc}}(r)|
            \sqrt{r^{2}+\left(\frac{\mathrm{d}r}{\mathrm{d}\phi}\right)^{\!2}}\mathrm{d}\phi
            \, ,
\end{equation}
where $\phi_{0}$ and $\phi_{1}$ are the entrance and exit angle, respectively.

{\bf Dynamical friction.} The nucleons approaching at a distance larger than $d_{\crm}$ from the center will not be captured but will still be deflected and hence take momentum from the PBH. This will also cause a force on the PBH, known as dynamical friction. In the NS reference frame, the change in momentum of a deflected neutron reads~\cite{Capela_2013}
\begin{equation}
    \Delta \bm{p}
        =
            \left(
                m_{n} v\.\gamma^{2}\.(\cos\theta - 1),\,
                m_{n} v \gamma \sin \theta,\,
                0
            \right)
            ,
\end{equation}
where $\theta$ is the total deflection experienced by the neutron in the PBH reference frame. Expressing the integral in terms of $x = R_{\rm s}/r$, the deflection angle is given by
\begin{equation}
    \theta(d)
        =
            -\pi+2 \frac{d}{R_{\rm s}} v\gamma \int_{0}^{x_{\max }} \frac{\mathrm{d} x}{\sqrt{\gamma^{2}-U(x)}}
            \, .
\end{equation}
All components of $\Delta p$ not parallel to the direction of motion of the PBH will cancel out by symmetry if one assumes that the density is approximately constant on the scales of interest. Adding a small correction by considering the relativistic momentum density for a perfect fluid to the result from Ref.~\cite{Capela_2013}, the force experienced by the PBH is given by
\begin{equation}
    \bm{F}_{\mathrm{DF}}
        =
            -2\pi v^{2}\gamma^{2}
            \left(
                \rho + P
            \right)
            \frac{\bm{v}}{v} \int_{d_{\crm}}^{d_{\rm max}} \beta\left\{1-\cos[\theta(\beta)]\right\}\mathrm{d}\beta
            \, .
\end{equation}
The upper bound of the integral $d_{\rm max}$ is defined as the distance at which the scattered nucleon no longer gains enough energy to be ejected from the Fermi sea~\cite{Génolini_Serpico_Tinyakov_2020}, this being equivalent to the local binding energy $\mu( r )$. Assuming that all the NS matter is composed by neutrons, the local value of $d_{\rm max}$ will be the impact parameter $d$ which solves the equation
\begin{equation}
    \mu
        =
            \sqrt{m_{n}^{2} + ( m_{n} v \gamma )^{2}
            \left\{
                [
                    1-\cos \theta(d)
                ]^{2}\.\gamma^{2}
                +
                \sin^{2}\theta(d)
            \right\}}
            -
            m_{n}
            \, ,
\end{equation}
where we have updated the equation given in Ref.~\cite{Capela_Pshirkov_Tinyakov_2013} with the appropriate relativistic effects. The total energy loss from dynamical friction is then
\begin{equation}
    \Delta E_{\mathrm{DF}}
        =
            \int_{\phi_{0}}^{\phi_{1}}|F_{\mathrm{DF}}(r)|\sqrt{r^{2}+\left(\frac{\mathrm{d}r}{\mathrm{d}\phi}\right)^{2}}\mathrm{d}\phi
            \, .
\end{equation}
For both DF and accretion the calculations have been made under the assumption that the gravitational effects of the PBH dominate the neutron's behaviour locally. Given that $d_{\rm max} < \mathcal{O}(10 R_{\rm s})$ in the core of the NS{\,---\,}where most of the energy loss from DF occurs{\,---\,}we expect this approximation to hold.

The dynamical friction and accretion calculations are done under the assumption that the neutrons in the PBH's wake are approximately collisionless. This is a valid assumption in the limit $c_{\rm s} \ll v$, which however starts to break down in the core of the NS, and in the case of the $2.12\,M_{\odot}$ NS does not hold at all since the sound speed is larger than the PBH velocity for most of the interior (although this is dependent on the chosen equation of state). 

Implementing a fully collisional treatment might significantly enhance the predicted energy loss from both channels. Previous analyses have shown that there is a strong enhancement of dynamical friction when $v \approx c_{\rm s}$, since a gravitational ``sonic boom'' creates a divergence in the force when the speed of the perturber is equal to the medium's speed of sound~\cite{Ostriker_1999}. This process has been extended to the case of relativistic dynamical friction in a collisional fluid~\cite{Barausse_2007} under the assumption of a weak gravitational field coming from the medium, which is not valid in the case of a NS, since the gravitational wake would be disrupted by the star's gravitational effect. We also stress that current collisional dynamical friction models are not suited to take into account the suppression coming from the binding energy of neutrons, which reduces the DF effects up to a factor of $10$ in our calculations.

%% file: Realistic.tex
The PBH energy loss in the NS interior is highly dependent on the local density, PBH velocity and binding energy per nucleon. Therefore, it is important to have an accurate description of the star's interior. 

We use the BSk-20 model~\cite{Potekhin_2013} to solve the TOV equations to find a numerical solution for the star interior; some of the relevant values are shown in Table~\ref{interior}. The Brussels-Montreal (BSk) models for energy-density functionals are a series of analytical models based on zero range effective interactions between nucleons~\cite{article}. These allow to obtain analytical approximations for the equation of state of a neutron star. We chose to work with theBSk-20 model since it has the best fit to the mass-radius data collected from the LIGO/VIRGO data from the GW170817 NS merger~\cite{Abbott_2018, Chamel_2016}.
\begin{figure}[t]
    \includegraphics[width=0.45\textwidth]{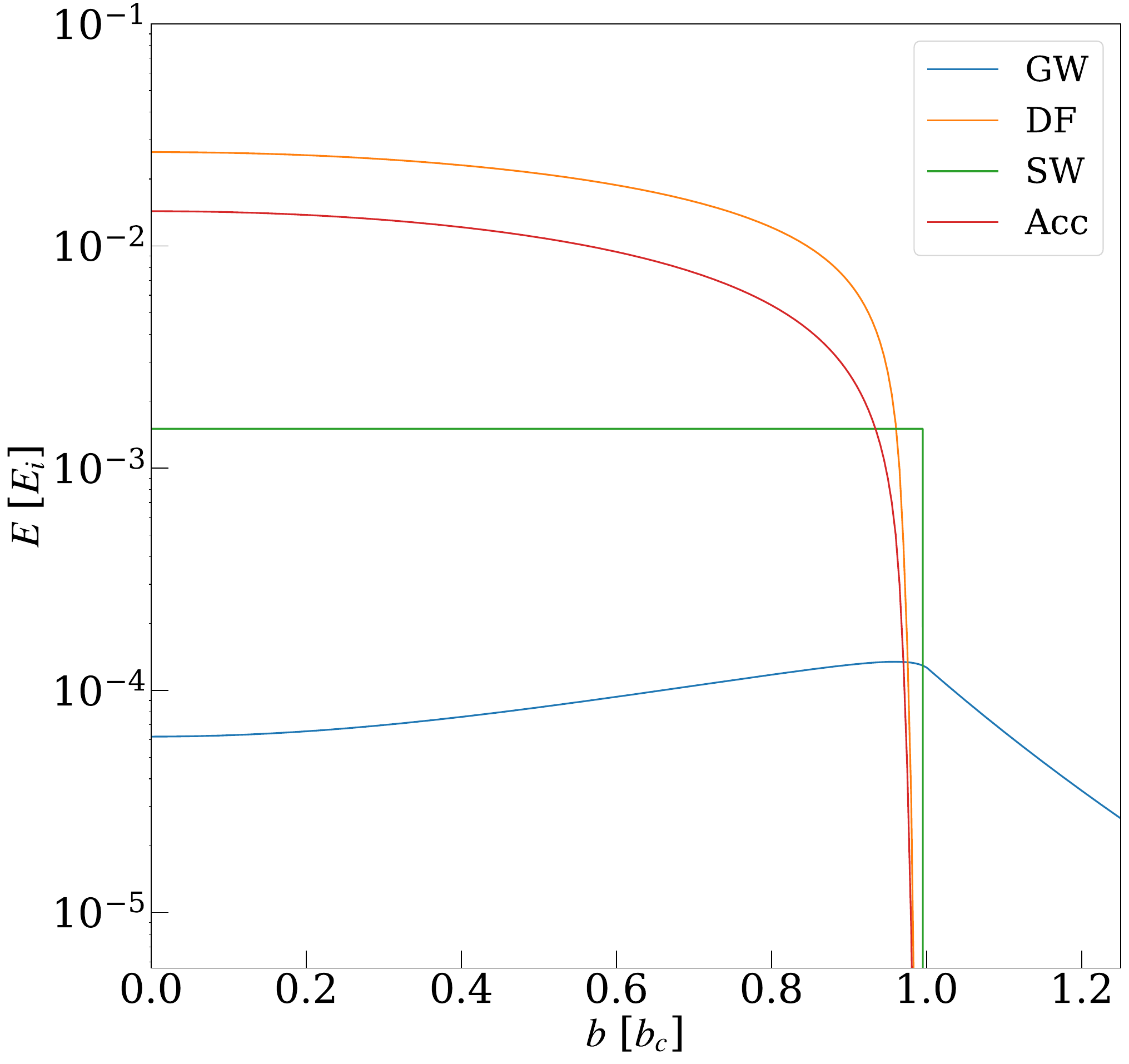}
    \caption{
        Total energy loss and energy loss for each mechanism versus impact parameter for a PBH of mass $10^{-9}\,M_{\odot}$ and initial velocity $10^{-3}$, and a NS of mass $1.52\,M_{\odot}$.
        }
\label{Energy_l}
\end{figure}

We use our interior solutions to numerically evaluate the Newtonian paths inside a NS. To these we superimpose the local GR velocities and interior values to compute the energy loss from dynamical friction and accretion on any given path. The GW energy loss in the interior is computed using the constant-density approximation as discussed in Section~\ref{sec:Section A}, since $\Delta E_{\mathrm{gw}}$ is subdominant to the other energy losses in the star's interior, this should not have a significant effect on the final results.

The energy loss from each component and the total energy loss are computed for two NS interiors: one which represents a NS of mass $1.52\,M_{\odot}$ (which is close to the expected mean mass for MSP) and one which represents an extremal $2.12\,M_{\odot}$ NS (which is close to the maximum allowed mass for MSP)~\cite{antoniadis2016millisecond}. The result for the $1.52\,M_{\odot}$ NS are shown in Fig.~\ref{Energy_l}.

The inclusion of a complete description enhances the dynamical friction and accretion for PBHs at low values of the impact parameter $b$ when passing through the star's core, but decreases the energy loss for PBHs moving only through the less dense crust around $b \approx b_{\crm}$. For a radial orbit and for a realistic interior, the result is a more then doubled in energy loss. Overall this results in a higher capture rate (by approximately~30\%) for a NS with realistic interior in environments with high velocity dispersion (like the GC).

The higher density and mass of the heavier star also have a strong impact on the total energy lost by the PBH, allowing the extremal star to capture more PBHs. The difference in energy loss reaches a factor of 6 for radial orbit, this together with a slightly higher critical impact parameter causes a significantly higher capture rate for the more massive star. 

We also run analyses for the BsK-19 and BsK-21 models~\cite{Potekhin_2013}, which represent the two extrema of our uncertainty on the mass-radius function for NS, representing denser and less dense NS models. Using these equations of state, we compute interior solutions for NS of the same mass ($1.52\,M_{\odot}$), to keep the main parameter for the PBH energy loss constant among the different models. As expected, denser stars have a higher maximal energy loss but a smaller critical impact parameter. While the velocity dispersion is small, the capture rate is dominated by GWs, and are hence  the same for the different equations of state (since the NS masses are the same), as the velocity dispersion increases, the effects from the changes in the critical impact parameter and maximal energy loss compete against each other, with the denser star leading to a higher capture rate when the velocity dispersion is high, similarly to the previously discussed cases. Overall, the associated changes in capture rate are at most $\sim 30\%$. 

\begin{table}[t]
    \centering
    \begin{tabular}{c|ccc}
\hline \hline 
& \text {$\rho_{0}^{(1)}=10^{18}\,\text{kg}/\text{m}^3\vphantom{1^{^{^{^{1}}}}}$} & \text { $\rho^{(2)}_{0} = 2\.\rho_{0}^{(1)}$ }  \\[1mm]
\hline 
$M$ [$M_{\odot}$] & 1.52 & 2.12$\vphantom{1^{^{^{^{1}}}}}$ \\[1mm]
$R$ [km]& 11.7 & 10.7 \\[1mm]
$\mu(r = 0)$ [MeV] & 271 & 812 \\[1mm]
$c_s(r = 0)$ [$c$] & 0.680 & 0.971 \\[1mm]
$v(r = 0)$ [$c$] & 0.797 & 0.939 \\[1mm]
$P(r = 0)$ $\big[ 10^{34}\,\text{N}/\text{m}^2 \big]$ & 1.57 & 8.11 \\[1mm]
\hline \hline
    \end{tabular}
    \caption{Relevant values for the two NS interior solutions.}
    \label{interior}
\end{table}

%% file: Central_parsec.tex
In order to analyze the PBH capture undergone by NSs in the GC one must have a good description of the dark matter density, stellar mass and velocity distribution within this region. In this section we collect these relevant pieces of information: stellar distribution, dark matter distribution and the expected velocity distribution of dark matter in the reference frame of local NSs.

\textbf{Stellar density.} The stellar density near the Galactic center can be modelled from high-resolution adaptive optics imaging~\cite{Genzel_2003}, which gives as the best-fit description of the central stellar cusp
\begin{equation}
    \rho_{\star}( r )
        =
            1.2 \times 10^6
            \left(
                \frac{r}{0.39\,\mathrm{pc}}
            \right)^{\!-\alpha}
            \,
                M_{\odot}\,\mathrm{pc}^{-3}
            ,
\end{equation}
where $r$ is the radial distance from Sagittarius A*. The function is split in two regimes with $\alpha = 1.4$ for $r < 0.39\,\mathrm{pc}$ and $\alpha = 2.0$ for $r \geq 0.39\,\mathrm{pc}$. From this one can derive an approximate stellar number density as $n_{\star}( r ) = \rho_{\star}( r ) / \langle M_{\star} \rangle$, where $\langle M_{\star} \rangle = 1.3\,M_{\odot}$ is the average mass for the Salpeter mass function.

\textbf{Pulsars age.} An important parameter for dynamical capture in pulsars is the age of the NS, given that likelihood of PBH capture is directly proportional to age of the NS. Regular isolated pulsars have an expected lifetime of $10\,\text{--}\,100\,{\rm Myr}$~\cite{Faucher_Giguere_2006}, meaning their expected age should be of order $\sim 10\,{\rm Myr}$ (although younger pulsars are easier to detect~\cite{Faucher_Giguere_2006}).

It is harder to directly estimate the age of MSPs given that the standard mechanisms. The characteristic age, does not work for MSPs, since they have received angular momentum from an external source~\cite{agec}. The expected age of a MSP in the GC can still be estimated by looking at other old stars in the GC, this process gives an expected age of $\langle T_{\rm GC} \rangle = 10.4\,\mathrm{Gyr}$~\cite{Crocker_2022}.
\begin{figure}[ht]
    \includegraphics[width=0.45\textwidth]{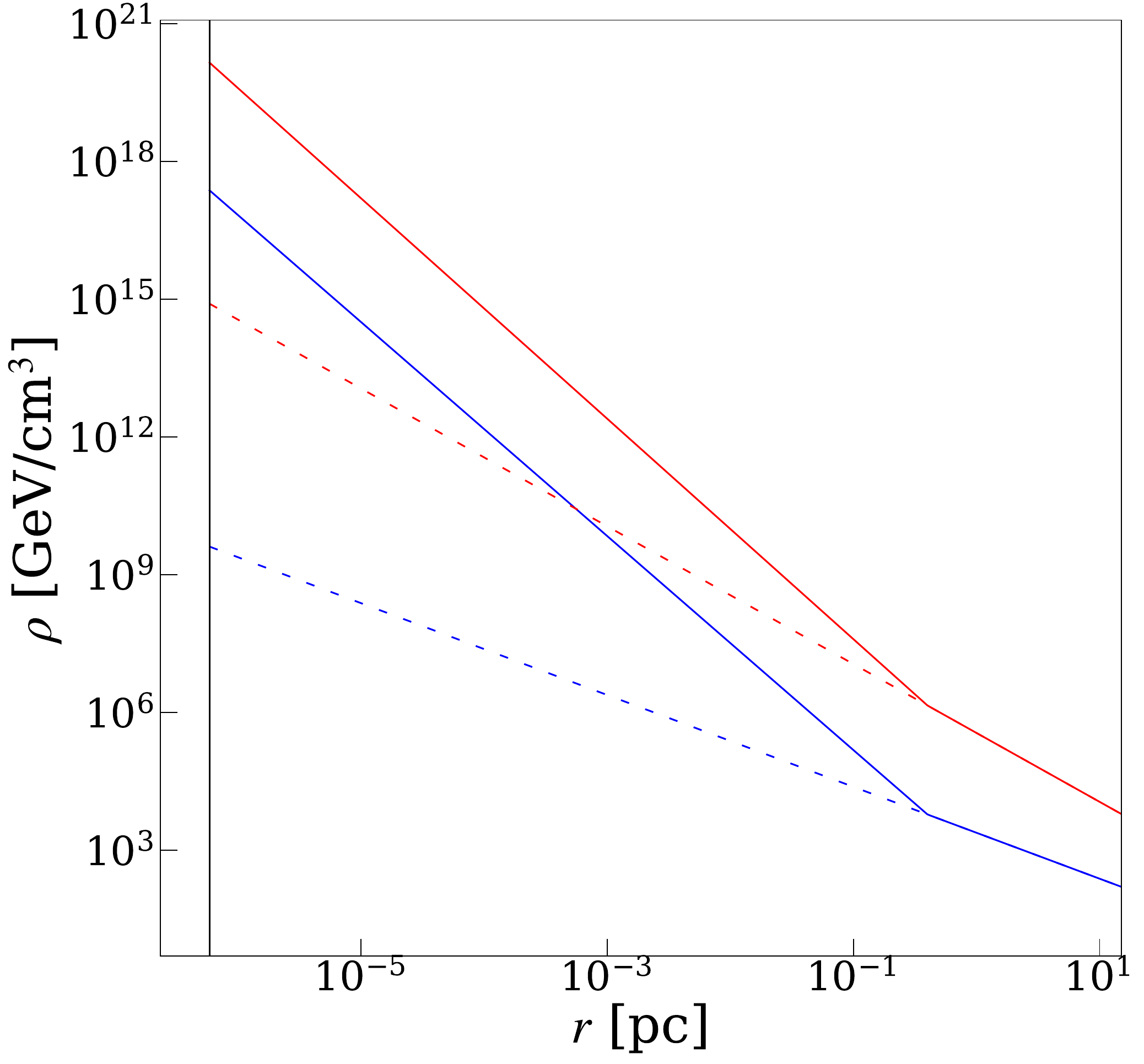}
    \caption{Dark matter density versus distance from Sagittarius A*, the black line represents the ISCO. Spiked (full lines), spikeless (dotted lines), NFW (blue) and Adiabatic (red) profiles are shown. See main text for further details.
    }
    \label{dm}
\end{figure}

\textbf{PBH density.} We assume that the radial PBH density and velocity profiles matches those of the dark matter. In the case where they compose only a fraction of the total dark matter, we simply rescale the density profile normalization. The dark matter distribution in the central parsec is unfortunately hard to infer from observations, since the gravitational potential is completely dominated by baryons. The best strategy to estimate the density profile is to extract it from numerical simulations which include baryonic effects, and which successfully reproduce the rotation curve and morphological properties of the Milky Way~\cite{Calore:2015oya, Schaller:2015mua, Bozorgnia:2016ogo}. Here, we consider two models for the dark matter density profile. The first is the standard Navarro-–Frenk–-White (NFW) model, which provides an excellent fit in dark matter simulations without baryons. The second is the “adiabatic-growth” model, labelled “A” below. The latter assumes that the baryons contracted adiabatically and symmetrically within the pre-existing dark matter halo, compressing the dark matter and increasing its density~\cite{Blumenthal1986}. When applied to an halo with an initial NFW profile, the result is a steeper profile inside the solar circle, with power-law index 1.5~\cite{Gnedin:2004cx}. For both the NFW and the adiabatic profiles, we considered the case with (subscript 's') and without (subscript 'n') a central `spike' induced by the adiabatic growth of the central black hole~\cite{Quinlan:1994ed, Gondolo:1999ef}, as described in Ref.~\cite{Bertone_2005} and shown in Fig.~\ref{dm}. The general form of the Galactic dark matter density inward of the solar system reads
\begin{equation}
    \rho_{\rm DM}
        =
            \begin{cases}
                \rho_{\odot}\left(\frac{R_{\odot}}{r} \right)^{\gamma_{\crm}},\;\;r_{\rm s}<r\leq R_{\odot}\\[2mm]
                \rho_{\odot}\left(\frac{R_{\odot}}{r_{\rm s}} \right)^{\gamma_{\crm}}\left(\frac{r_{\rm s}}{r} \right)^{\gamma_{\rm s}},\;\;r_{\rm in}<r\leq r_{\rm s}
                \, ,
        \end{cases}
\end{equation}
where $\rho_{\odot}$ represents the local density in the solar system, $\gamma_{\crm}$ is the power-law index for the cusp, $r_{\rm s} = 0.4\;\text{pc}$ is the distance at which the DM spike behaviour begins, $r_{\rm in} = 8 \times 1 0^{-7}\,$pc is the smallest distance at which a particle can orbit the black hole, and $\gamma_{\rm s}$ is the power-law index for the spike. For the non-spiked case $\gamma_{\rm s} = \gamma_{\crm}$. The relevant values for all models are shown in Table~\ref{models}.
\begin{table}[ht]
    \centering
    \begin{tabular}{c|cccc}
\hline \hline 
 &\text {NFW,n} & \text {NFW,s}& \text {A,n} & \text {A,s}  \\
\hline 
$\rho_{\odot}$ [GeV/cm$^{3}$] & 0.3 &0.3 &0.5&0.5 \\
$\gamma_{\crm}$ & 1 & 1& 1.5& 1.5 \\
$\gamma_{\rm s}$& 1 & 2.33& 1.5& 2.4 \\[0.5mm]
\hline \hline
    \end{tabular}
    \caption{Relevant values for the four dark matter models.}
    \label{models}
\end{table}

\textbf{PBH velocity distribution.} The velocity distribution for a spherically-symmetric isothermal system, as is the dark matter halo, should approximately follow a Maxwellian distribution~\cite{Lisanti_2016}. From the virial theorem, one has
\begin{equation}
    v_{\rm rms}
        =
            \sqrt{\frac{GM_{\crm}}{r}}
            \, ,
\end{equation}
where $M_{\crm}( r )$ is the total mass contained in the sphere of radius $r$ centered at Sagittarius A*, and $v_{\rm rms}$ is the root-mean-square velocity. $M_{\crm}$ is obtained by integrating the stellar-DM density up to a sphere of radius $r$, and then adding the mass of Sagittarius A* mass, $M_{Sag} = 4.1\times10^6 M_{\odot}$~\cite{2019} as the central-mass value. The velocity distribution can then be written as
\begin{equation}
    f( \bm{v} )
        =
            \left(
                 \frac{3}{2\pi}\.\frac{1}{v_{\rm rms}^{2}}
            \right)^{\!3 / 2}
            \exp\!
            \left(
                - \frac{3}{2}\.\frac{v^{2}}{v_{\rm rms}^{2}}
            \right)
            .
            \label{eq:maxwell}
\end{equation}
The approximate velocity distribution obtained in this way gives results close to the observed velocity dispersion near the galactic center~\cite{Sch_del_2009}, although the data has a slight suppression of radial velocity, likely due to interaction with the dense environment and with the supermassive black hole~\cite{Genzel_2000}. This bias might be larger at closer distances from Sagittarius A* but in the distances considered in this paper this should at most add $\mathcal{O}(10\%)$ corrections to the velocity distribution, and more likely negligible corrections if $v_{\rm rms}$ is not substantially changed. There is little difference in expected velocity between different dark matter distribution models, especially in the central parsec, which is of interest here. For $r < 0.1\,$pc, the velocity dispersion is dominated by the supermassive black hole mass such that $v_{\rm rms} \propto r^{-1/2}$.

Equation~\eqref{eq:maxwell} describes the velocity distribution of dark matter for a static observer and is the one usually used in most of the literature on dynamical capture. However, any neutron star in the central parsec will also be orbiting around Sagittarius A*. We calculate the velocity distribution by moving to the frame of reference of a NS moving with velocity $v_{\rm NS} = v_{\rm rms}$ through the dark matter halo. This is the case for a NS moving on a circular orbit in the GC, which gives the velocity distribution
\begin{equation}
    f( v )
        =
            \sqrt{\frac{6}{\pi}} \frac{v} {v_{\rm rms}^{2}} \frac{\sinh(3 v / v_{\rm rms})}{\exp\!\left(3 \big[ v^{2} + v_{\rm rms}^{2} \big] / 2 v_{\rm rms}^{2} \right)}
            \, .
\end{equation}
This velocity distribution reduces the capture rate by a factor 4.5 with respect to the standard Maxwell distribution in the small-$v / v_{\rm rms}$ limit, which is the relevant regime for low-mass PBH capture.

%% file: Capture.tex
\begin{figure}[t]
    \includegraphics[width=0.45\textwidth]{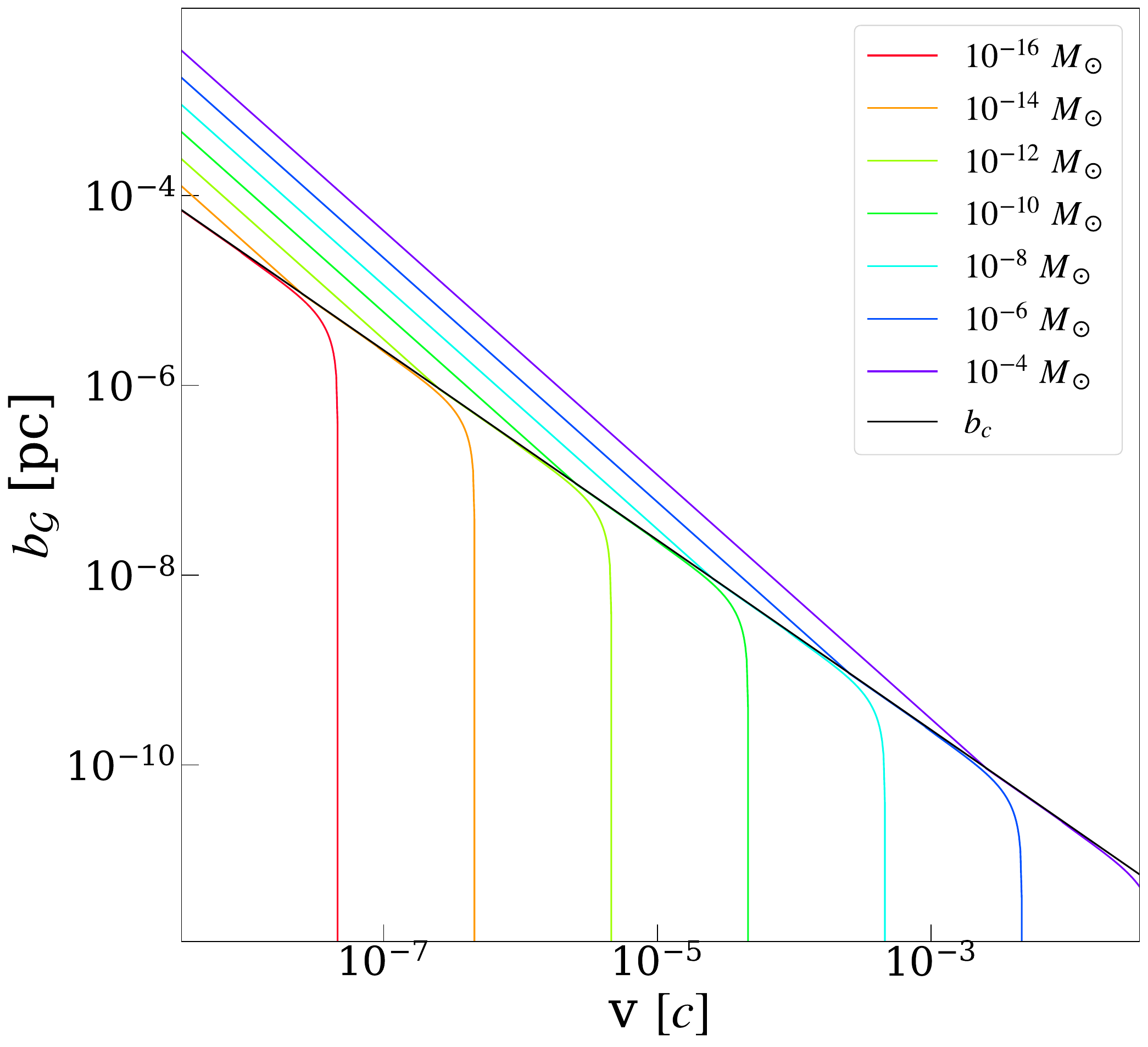}
    \caption{Capture radius $b_{\Gcal}$ as a function of PBH's initial speed for a NS of mass 1.52$\,M_\odot$.}
    \label{cr}
\end{figure}
Assuming a locally-uniform PBH distribution, it is possible to calculate the NS capture rate $\Gcal$, which is the rate at which PBHs are captured into a bound orbit at any point in the Galaxy~\cite{Génolini_Serpico_Tinyakov_2020}
\begin{equation}
    \Gcal
        =
            \int\mathrm{d}{v}\;
            4\pi v^{3} \.n_{\rm PBH}
            f( v )\.\pi\.
            b_{\Gcal}^{2}
            \, ,
\end{equation}
where $b_{\Gcal}( v )$ is the maximum radius at which capture can occur for a given velocity $v$ (see Fig.~\ref{cr} for numerical evaluation), its distribution $f( v )$ (in the NS reference frame), and the local number density of the PBHs $n_{\rm PBH} = \rho_{\rm PBH}/m$.

\begin{figure}[t]
    \includegraphics[width=0.45\textwidth]{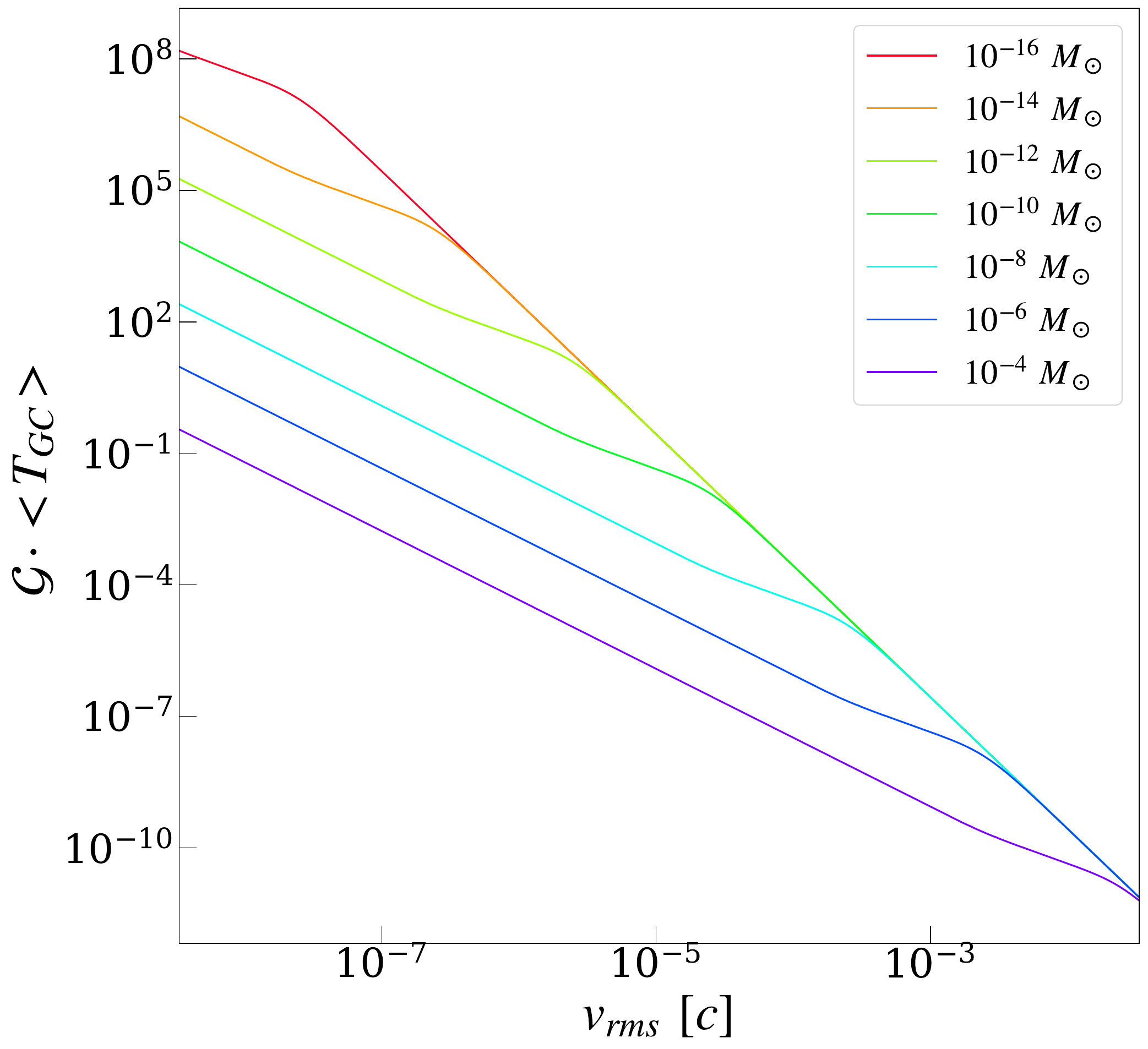}
    \caption{Expected number of captures as a function of the local $v_{\rm rms}$, with local PBH density of 1 GeV/cm$^{3}$, for a MSP of mass 1.52$M_\odot$ and age $<T_{\rm GC}>$.}
    \label{cap}
\end{figure}

The capture rate might be enhanced by a factor of 3.5 for tight binaries according to numerical simulations~\cite{Brayeur_2012}. This is relevant for MSPs which are expected to be in tight binaries for a large portion of their existence, thereby accelerate their angular velocities up to the observed high values~\cite{Tauris_2011}. Including this effect, it is then possible, for a MSP moving through a circular orbit in the Galactic center, and using the velocity distribution discussed before, to plot the capture rate/velocity-dispersion graph for different masses as shown in Fig.~\ref{cap}. The PBH capture rate is systematically enhanced for the heavier neutron star ($2.12\,M_\odot$), where the difference is larger (up to a factor of five) for higher velocity dispersions, since the larger maximum possible capture velocity $v_{\rm max}=\sqrt{2\Delta E_{\rm max}/m}$ of the denser star becomes more relevant to the capture statistics in such environments. 

The capture rates for different masses tend to converge at high velocities. This can be explained by analysing the behaviour of the capture radius-velocity relation as given in Fig.~\ref{cr}. For all masses, $b_{\Gcal}$ converges to the critical impact parameter before rapidly collapsing to zero as the maximum possible energy loss of a PBH passing through a radial trajectory, $\Delta E_{\rm max} = \Delta E( b = 0 )$, becomes insufficient to disperse the original kinetic energy of the PBH. This first convergence is clearly caused by the sudden jump in energy loss as the PBH crosses the NS's surface, which means that, for velocities close to $v_{\rm max}$, the capture radius is approximately constant and similar to $b_{\crm}$.

In high-velocity-dispersion environments like the GC, most of $v_{\rm max}$ for different masses are concentrated at the beginning of the velocity distribution where the velocity is much smaller than the velocity dispersion. This property extends to more masses the higher the value of $v_{\rm rms}$. Expanding around $v \ll v_{\rm rms}$ to first order, one can approximate the velocity distribution of PBHs as 
\begin{equation}
    f( v )
        \approx 
            3\.\sqrt{\frac{6}{e^{3}\pi}} \frac{v^{2}}{v_{\rm rms}^{3}}
    \propto
            v^{2}
            \, .
\end{equation}
The same proportionality holds also for the regular Maxwell distribution discussed in the literature. The largest contribution to $\Gcal$ comes from velocities close to $v_{\rm max}$ where $b_{\Gcal}\approx b_{\crm} \approx R\.v_{\star}/v$, meaning the integrand is roughly proportional to $v$. The capture rate will then be proportional to
\begin{equation}
\begin{aligned}
    \Gcal
        \propto
            \int_{0}^{v_{\rm max}}\d v\;
            n_{\rm PBH}\.v
        \propto
            \frac{\rho_{\rm PBH}}{m}\.v_{\rm max}^{2}
            \\[1.5mm]
        \propto
            m^{-1}\.\frac{\Delta E_{\rm max}}{m}\,
        \propto\,
            \text{constant}
            \, ,
\end{aligned}
\end{equation}
where in the last step, one needs to use the property $\Delta E \propto m^{2}$ discussed in Appendix~\ref{B}. Therefore, the convergence of the capture rate is a general feature of the capture dynamics for PBHs with Maxwell-like velocity distributions, and it is more pronounced in environments with larger velocity dispersions. As a consequence of this convergence, the final PBH abundance constraints exhibit a flat behaviour as a function of mass.

%% file: Time.tex
Once the PBH is captured in a bound orbit with a NS, it will take some time for the PBH to lose enough energy to fall to the center of the star and consume it. If the time required for the PBH to destroy the star is longer than the age of the star, then the capture might not yield any observable consequence and such captures must be ignored in the calculations.

Immediately after capture, the PBH will be in an elliptical orbit of energy 
\begin{equation}
     E_{\mathrm{orb}}
        =
            -
            \frac{GmM}{2a_{0}}
        =
            \frac{1}{2}\.m\.v_{i}^{2}
            -
            \Delta E_{\rm tot}
            \, .
            \label{semim}
\end{equation}
where $a_{0}$ is the semi-major axis of the initial captured orbit, and $\Delta E_{\rm tot}$ is the total energy lost by the PBH by summing over all the energy loss channels from Section~\ref{sec:Section-B}. Assuming that the orbit is such that the distance of closest approach $r_{\prm}$ is the same~\cite{Génolini_Serpico_Tinyakov_2020}, and approximating the approaching orbit as nearly parabolic, yields $b_{0} \approx b \sqrt{a_{0} / a}$. The collapse of the orbit can then be split in two cases: the GW-induced collapse for $b > b_{\crm}$. and the collapse dominated by the energy lost inside the NS for $b \leq b_{\crm}$.

In the first case ($b > b_{\crm}$), the orbital decay will be completely dominated by the GW energy loss. The decay time for a highly eccentric binary is given by~\cite{PhysRev.131.435}
\begin{align}
\begin{split}
    T_{\rm decay}
        =
            \frac{15}{304M^{2}m}
            \left(
                \frac{a_{0}\left(1-e^{2}
            \right)}{1.135}\right)^{4}
            \\[1mm]
            \times \int_{0}^{e_{0}}\d e\;
            \frac{e^{-\frac{67}{19}}
            \left[
                1 + \frac{121}{304}\. e^{e^{2}}
            \right]^{1187/2299}}{\left(1-e^{2}\right)^{3 / 2}}
            \, ,
\end{split}
\end{align}
which can be approximated as
\begin{equation}
    T_{\rm decay}
        \approx
            3.53 \times 10^{-2}
            \frac{v_{i}^{7}\.b^{7}}{\sqrt{G^{5}\.m\.M^{3}R^{7}\.v_{\star}^{14}\.(-2\.E_{\mathrm{orb}})}}
            \, .
\end{equation}
If the energy lost by the PBH through gravitational waves is much larger than its initial energy, one can utilise Eq.~\eqref{GW_loss} such that $E_{\mathrm{orb}} \approx - \Delta E_{\text{GW}}$. Under the assumption $v_{\star} \gg v_{i}$, this yields
\begin{align}
\begin{split}
    T_{\rm decay}
        \approx&
            \,5.58\.\text{Myr}\left(\frac{b}{b_{\crm}}\frac{0.44}{v_{\star}}\right)^{\!21/2}\\[1mm]
        &\times
            \left(\frac{10^{-11} M_{\odot}}{m}\right)^{\!3/2}\!\left(\frac{M}{1.52\. M_{\odot}}\right)^{\!5/2}
            \, ,
\end{split}
\end{align}
which is similar to the result presented in Ref.~\cite{Génolini_Serpico_Tinyakov_2020} but with slightly different exponents.

In the second case ($b \leq b_{\crm}$), the energy loss will be dominated by the extremely fast intervals when the PBH crosses through the NS. Under the assumption that the first few orbits after capture follow a similar path through the neutron star as the initial hyperbolic orbit (which should be the case given the small effect on the kinetic energy of the PBH while inside the NS), the energy loss for each orbit around the NS will be approximately equivalent. One can then approximate the total decay time as the sum of the orbital periods $T_{ell} = 2 \pi \sqrt{ a_n^{3} / GM}$ that the PBH goes through as the orbit decays. This gives
\begin{equation}
    T_{\rm decay}
        \approx
            2\pi GM
            \left(
                \frac{m}{2\Delta E_{\rm tot}}
            \right)^{\!3/2}
            \zeta\!
            \left(
                \frac{3}{2},-\frac{E_{\mathrm{orb}}}{\Delta E_{\rm tot}}
            \right)
            ,
\end{equation}
where $\zeta$ is the Hurwitz zeta function. $T_{\rm decay}$ is roughly proportional to $m^{-3/2}$, and will therefore be considerably higher for lower masses. This puts constraints on the minimum PBH mass which can effectively disrupt a NS within the total age of the star. For an old MSP of age $10\,$Gyr and mass $1.52\,M_{\odot}$, the minimum mass able to cause a collapse is $m_{min} \approx 10^{-16} M_{\odot}$. Once the orbit has completely decayed, the mass accretion rate of the black hole can be modeled by the spherical Bondi accretion, which gives as the time for the NS to collapse into a black hole~\cite{Génolini_Serpico_Tinyakov_2020}
\begin{equation}
    T_{\rm collapse}
        =
            \frac{c_s^{3}}{4 \pi \lambda \rho\.G^{2}\.m}\approx 1.35\,\text{yr}
            \left(
                \frac{m}{10^{-11} M_{\odot}}
            \right)
            .
\end{equation}
In the last stages of stellar collapse, the fast rotation of the NS will slow down the accretion and the new regime will be Eddington-like. However, this will be relevant only in the last moments of the star's collapse~\cite{Kouvaris_2014}. This does not impact our analysis on the scale of pulsar lifetimes. $T_{\rm collapse}$ is then an order of magnitudes smaller than $T_{\rm decay}$ for all PBH masses of interest and can therefore be ignored.

%% file: Collapses.tex
\begin{figure}[ht]
    \includegraphics[width=0.43\textwidth]{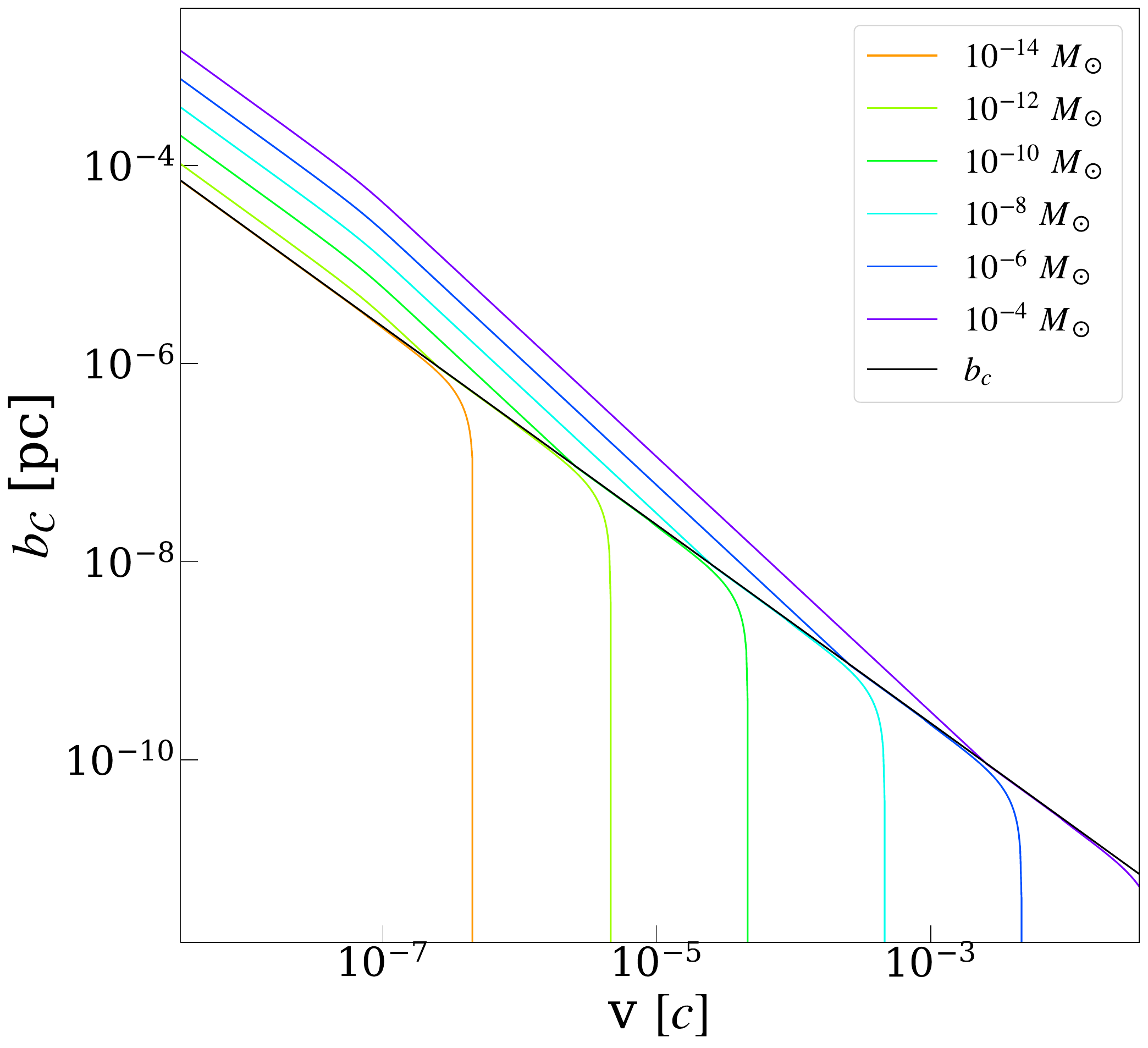}
    \caption{Critical radius for collapse as function of PBH initial velocity for a NS of mass 1.52\,$M_{\odot}$ and age $\langle T_{\rm GC} \rangle$.
    \vs{-2mm}}
    \label{cr2}
\end{figure}
Given a NS of age of $T_{\star}$, it is possible to find the expected number of captures leading to collapse within its lifetime. We call this the expected collapse number $\Ccal$. Of course, no star can undergo more than one collapse, but this number just represents the average number of PBH encounters capable of leading to collapse that the star should have undergone in its lifetime. 

\begin{figure}[t]
    \includegraphics[width=0.43\textwidth]{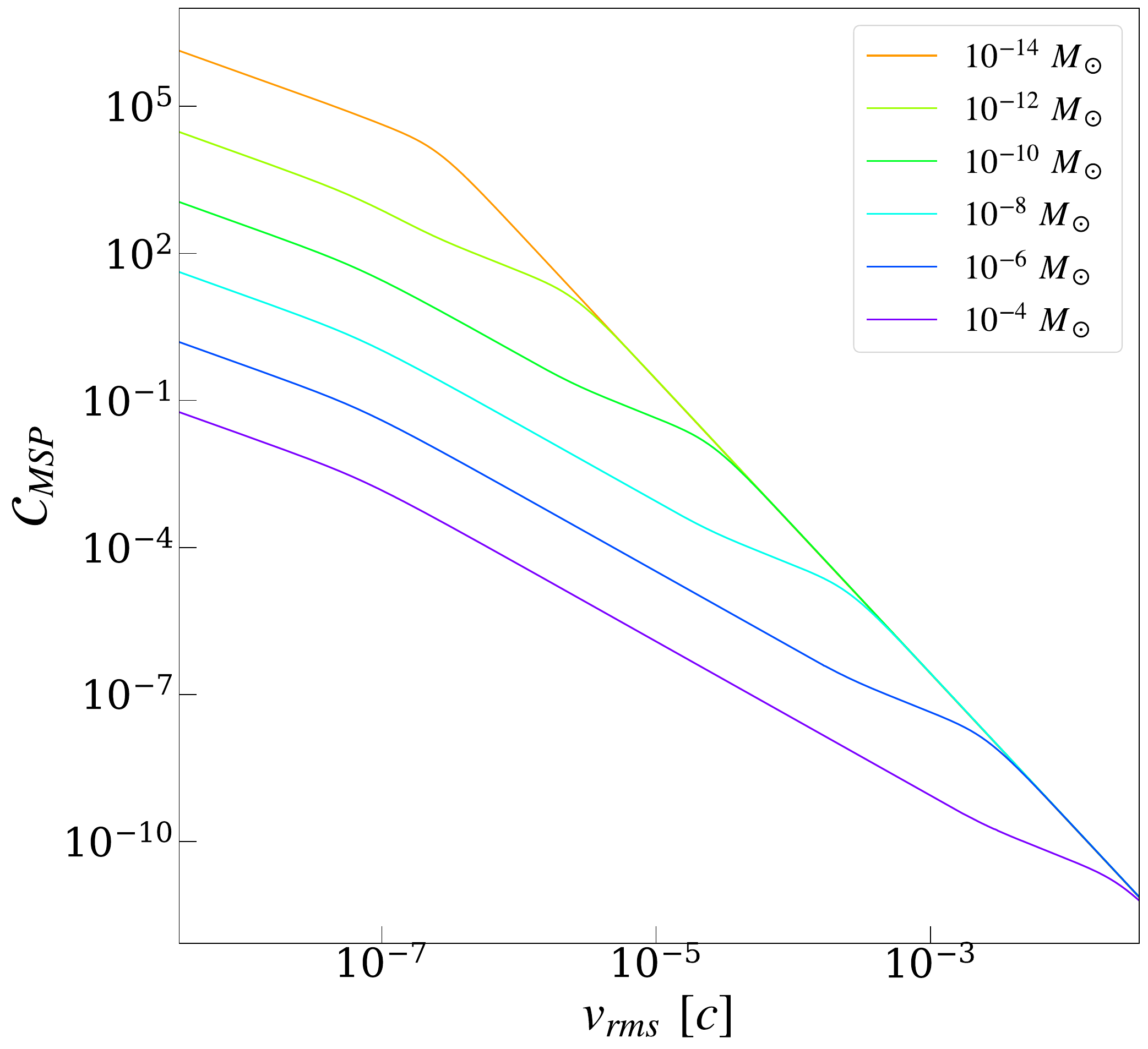}
    \caption{
        Expected number of collapses as a function of the local $v_{rms}$, with local PBH density of $1\,$GeV/cm$^3$, for a MSP of mass 1.52$\,M_{\odot}$ and age $\langle T_{\rm GC} \rangle$.
        }
    \label{coll}
\end{figure}

We define the critical radius $b_{\Ccal}(m,v)$ for collapse, as the maximum distance such that a PBH of mass $m$ and velocity $v$ is captured and falls into the NS within a time $T < T_{\star}$. PBH masses for which any capture orbit takes more than the age of the MSPs to collapse will have $b_{\Ccal} = 0$ for any velocity. As an example in Figs.~\ref{cr2} and~\ref{coll}, PBHs with mass $m_{\rm min} \approx 10^{-16}\,M_{\odot}$ or smaller cannot cause a collapse within the age of the star. Hence, they have $b_{\Ccal} = \Ccal = 0$ and do not appear in the log-log graph. 

The orbital decay time $T_{\rm decay}( b, v )$ can be comparable to the star's life. In that case, there is a shorter time window for capture and collapse $T_{\rm eff} = T_{\star} - T_{\rm decay}( b, v )$. In order to account for this, one can start by defining an effective area of capture for a PBH with mass $m$ and velocity $v$ as
\begin{equation}
    A_{\rm eff}
        =
            \int_{0}^{b_{\Ccal}}(T_{\star} - T_{\rm decay})2\pi r \;\mathrm{d}r
            \, .
\end{equation}
This gives the expected collapse number,
\begin{equation}
    \Ccal
        =
            \int n f({v}) \,8\pi^2 v^3 \int_{0}^{b_{\Ccal}}
            \left(
                T_{\star} - T_{\rm decay}
                \right)
                 r \;\mathrm{d}r\.
            \mathrm{d} v
            \, .
\end{equation}
In the following calculations, we approximate the decay time as the radial orbit decay time $T_{\drm}$, such that $T_{\drm} = T_{\rm decay}( b = 0, v )$, which is a good approximation in high-velocity-dispersion environments for masses much larger than $m_{\rm min}$. The time dampening is strongest for lower mass PBHs with larger orbits.
\newpage

An important question is if, in the case of MSPs, the companion star might negatively impact the orbital collapse. It is possible that the PBH might be slingshot out of the system through the gravitational interaction with the other two bodies. Numerical simulations are necessary to check the actual effect on the expected collapse number.

The expected number of collapses $\Ccal$ for a given NS can be interpreted as the mean number of PBHs, with the correct characteristics leading to the star's collapse, which should have crossed the NS in its lifetime. Given that PBH captures are independent of each other, the likelihood for a NS to have encountered a given number of PBHs with the correct mass, velocity and impact parameter to lead to a collapse will be given by a Poisson distribution with mean $\Ccal$. The probability of the NS's survival will then be given by $P( \text{Collapse} ) = 1 - e^{-\Ccal}$. In general the expected suppression rate of an initial population of $N_{0}$ neutron stars at a distance $r$ from the GC after a time $t$ will be in the form 
\begin{equation}
    N( t )
        =
            N_{0} e^{-\Ccal(r,t)}
            \, .
\end{equation}
Since $\Ccal$ is roughly proportional to $t$, the suppression will be much higher for older populations of NSs.

We find that even assuming an adiabatic and spiked dark matter halo, a NS mass of $1.52\,M_{\odot}$ (which is larger than the average mass of regular pulsars~\cite{Zhang_2011}), and ignoring dense environment issues which will be discussed later in this section, the survival rate after the average pulsar lifetime is $N(10^7 \rm{yr}) \approx N_{0}$. This holds even at distances of $10^{-3}\,$pc, meaning it is impossible for the PBHs to disrupt a significant fraction of pulsars at galactocentric distances larger than that. This clearly cannot explain the lack of pulsars detection in the inner $1\,\text{--}\,10$ parsecs~\cite{Dexter_2014}.

This means that dynamical capture of PBHs cannot be the dominant channel to explain the missing pulsar problem, unless other processes significantly increase the PBH capture rate (e.g.~capture from adiabatic contraction see Ref.~\cite{Capela_2013-2} and more recently in Ref.~\cite{10.1093/mnras/stae147}).

%% file: Constraints.tex
As mentioned in Ref.~\cite{Montero_Camacho_2019}, the assumption that the PBH-neutron star system is gravitationally isolated from the surrounding distribution of stars breaks down in dense stellar environments. The apoastron $r_{a}$ for a PBH of mass $m$, initial velocity $v$ and impact parameter $b$ is given by
\begin{equation}
    r_{a}
        =
            a_{0}
            \left(1+\sqrt{1-\frac{b^2v_{\star} ^2}{a_{0}\.GM}}\.\right)
            ,
\end{equation}
where $a_{0}$ is the semi-major axis of the initial bound orbit given by Eq.~\eqref{semim}. By approximating the local stars as uniformly distributed, we get an expected distance $d_{\star} \approx 1/n_{\star}^{1/3}$, which gives us $r_{\star}\approx d_{\star}\.\sqrt{M_{\rrm}} / ( \sqrt{M_{\rrm}} + 1 )$ as the cutoff of the gravitational influence of the NS, where $M_{\rrm}= \langle M_{\star} \rangle/M$. In the case of the central few parsecs, it is important to consider the pull of the central mass when compared to the NS at the PBH's apoastron. The distance at which the pull of the NS and the central mass are equivalent is $r_{\crm} = r \sqrt{M / M_{\crm}}$. We can then impose the condition $r_{a} \leq \min( r_{\star}, r_{\crm} )$ to all captures to make sure our two-body-orbital-collapse formalism is a valid approximation. In practice, at the Galactic centre we always find $r_{\crm} < r_{\star}$. Redefining $b_{\Ccal}$ to include this property for the bound orbit, allows us to re-evaluate it numerically for different radii. This turns out to exclude collapses for a wide range of low masses. The closer a NS gets to the GC, the more masses are excluded from capture due to the gravitational effects of the central mass. 

Assuming a Dirac delta for the mass function of PBHs, we set for a given PBH mass an upper limit on the PBH abundance, such that the likelihood of such an observation to occur for the given PBH population is under 5\%. We consider the possibility of setting such constraints from future observations of one or more MSPs in the central parsec. As mentioned before, we take the age of these objects to be $T_{\rm MSP} = 10.4\,$Gyr. 

In the case of a single observation it is straightforward to use $\Ccal \propto \rho_{\rm PBH}$ to find $f_{\rm PBH} = \rho_{\rm PBH} / \rho_{\rm DM}$, which would give a likelihood of survival for the MSP under 5\%,
\begin{equation}
    f_{\rm PBH}( m )
        \geq
            -\frac{\ln{0.05}}{\Ccal_{\rm DM}}\approx\frac{3}{\Ccal_{\rm DM}}
            \, ,
\end{equation}
where $\Ccal_{\rm DM}$ is the expected collapse number for $\rho_{\rm PBH}  = \rho_{\rm DM}$. Using this, we can plot the constraints on different PBH masses from the detection of a MSP at different distances from the GC, for different DM distributions and for MSPs of different masses. The possible scenarios for constraints from a single MSP observation are shown in Fig.~\ref{all}. 

Constraints above the upper limit $f_{\rm PBH}( m ) = 1$ are for unphysical scenarios in which the PBH abundance exceeds the DM abundance in the Milky Way. No physical constraint can be put from the observation of a single MSP in the NFW case, with or without a spike. However, as discussed before it seems that the adiabatic case should be the more likely physical scenario, in which case some constraints might be possible, especially in the presence of a spike around Sagittarius A*. There is a substantial enhancement of constraints from the observation of an extremal $2.12\,M_{\odot}$ MSP. In the adiabatic-spik case both an observation of an average mass or an extremal MSPs are enough to put physical constraints, while in the adiabatic non-spiked case only the observation of an extremal MSP would be enough. The most optimistic case for constraints from a single observation are shown together with previous constraints in Fig.~\ref{opt}.

The possibility to set constraints with future observations of pulsars located very close to the Galactic center depends on the detectability of these objects despite the strong inter-stellar scattering expected in this region~\cite{1998ApJ...505..715L}. Under our approximations, the above constraints effectively vanish in the currently unconstrained ``asteroid mass" window $( 10^{-16}\,\text{--}\,10^{-10}\,M_{\odot})$, due to the fact that Sagittarius A* and the dense stellar environment disrupt the orbits of captured PBHs at lower masses. It might however be possible to constrain this mass range by observing an old MSP in an ultra-faint dwarf with similar dark matter density and velocity dispersion but with lower baryon density. However, this does not seem to be a promising avenue both because we don't expect to see many pulsars in these regions, due to the low baryonic fraction. Furthermore, even if we could observe one, it would be hard to reconstruct the distance from the center of the dwarf with high precision.

In the future, it may become possible to set interesting constraints in this mass region, by estimating the global evolution of the PBH phase space due to interactions with the global compact-star population in the central parsec, and three-body gravitational interactions with other stars, in the gravitational potential of Sagittarius A*, and  how this may impact PBH capture in the region.
\vfill

%% file: conclusion.tex
In this work we have studied PBH-NS interactions and orbital dynamics. We have improved previous energy-loss calculations by treating the interaction of PBHs with NS using a realistic solution for its interior, and refining the treatment of the interaction dynamics and collapse likelihood.
\newpage

We found that dynamical capture of PBHs cannot explain the missing pulsar problem. The velocity dispersion is simply too high for the PBH to efficiently disrupt the star within the pulsar's lifetime ($\sim 10\,$Myr), even under the most optimistic assumptions. For the lower masses of the PBH's unconstrained region there are still an average of $10^{3}$ PBH-NS crossing in the pulsar lifetime at $10\,$pc from the GC, with more crossings closer to the GC. We have also shown that it is possible to set stringent constraints on sub-solar mass PBHs from MSP observations in the Galactic center. The main uncertainty in the calculation arises from the many-body dynamics of the central parsec. This affects the dynamics and capture of PBHs, including hyperbolic encounters with other stars and the effect of binary companions. These are expected to be present, allowing for the formation of MSPs, on the PBH's orbital collapse. It is important to investigate how all these processes will impact the orbital collapse of the captured PBHs. More constraints might become possible with a better understanding of the interactions between PBHs and the global population of pulsars at the Galactic center. 

These constraints would be significantly enhanced if multiple MSPs were observed near the Galactic center. This would be indicative of the presence of a large population of objects in a region with large dark matter density. We discuss for completeness the case of multiple pulsars detection in Appendix~\ref{A}.

In our analysis, both the dynamical friction and accretion calculations are performed under the assumption that the neutrons in the PBH's wake are approximately collisionless. We have argued that a fully collisional treatment might significantly enhance the energy loss from both channels. A detailed investigation of the collisional effects would likely strengthen the constraints derived in this paper and extend them to lower PBH masses. 
\vfil

%% file: appendix.tex
 We discuss in this appendix the implications of observing a population of pulsars. It is in fact possible that more than one pulsar will be observed within a given radius. And even if a single pulsar is detected, others likely exist whose poles are not pointing at the Earth and are thus undetectable ($\approx 8$ times the amount of detectable pulsars~\cite{Maciesiak_2012}).

If we observe $n$ pulsars and assume them to be the entire original population, and if we further assume stars to be on a shell at radius $r$ -- which is a suitable approximation given that larger radii will contain most of the stars within a given sphere around the GC -- then the likelihood of $n$ stars surviving is $P( s = n ) = e^{-n\.\Crm( r )}$. This is clearly an oversimplification, as the PBHs could have destroyed some but not all of the NS at a given radius. If we instead consider an expected number of neutron stars $N$ on shell at a radius $r$, we get the likelihood for $n$ or more survivals as 
\begin{equation}
    P( s \geq n)
        =
            I_{\exp(-\Crm)}(n, 1 + N - n)
            \, ,
\end{equation}
where $I_{\ldots}(\ldots, \ldots)$ is the regularized incomplete beta function. 

If a population of $n$ MSPs is found on a shell at a distance $r$, the na{\"i}ve constraints will take the form
\begin{equation}
    f_{\rm PBH}( m )
        \geq
            \frac{\ln{20}}{n\.\Crm_{\rm DM}( r )}
            \, ,
\end{equation}
so that the constraints are simply scaled by a factor of $n$. If instead we assume an initial population number $N$, the constraints for the observation of $n$ or more average-mass MSPs becomes
\begin{equation}
    f_{\rm PBH}( m )
        \geq
            - \frac{\ln{I_{0.05}^{-1}\left(n, 1+N-n\right)}}{\Crm_{\rm DM}( r )}
            \, ,
\end{equation}
where $I_{x}^{-1}( b, c ) = y$ is the inverse of the regularized incomplete beta function which solves the equation $I_{y}( b, c ) = x$.

%% file: appendix2.tex
In this appendix we show a useful transformation to convert the energy loss for given PBH mass and initial velocity to any other mass and initial velocity. First we need to make explicit the dependence on mass of the energy loss. We start by defining a re-scaled impact parameter $\tilde{d} = d / R_{\srm}$ to show explicitly that $\theta( \tilde{d} )$ is independent from the PBH mass:
\begin{equation}
    \theta(\tilde{d})
        =
            -\pi+2 \tilde{d} v\gamma \int_{0}^{x_{\max }} \frac{\mathrm{d} x}{\sqrt{\gamma^{2}-\left(1+\tilde{b}^{2}v^{2}\gamma^{2} x^{2}\right)(1-x)}}
            \, .
\end{equation}
We can then write explicitly the dependence of $\bm{F}_{\rm DF}$ on the PBH mass $m$. Since $\theta( \tilde{d} )$ does not depend on $m$, $\tilde{d}_{\rm max} = d_{\rm max}/R_{\srm}$ and $\tilde{d}_{\crm} = d_{\crm}/R_{\srm}$ are also independent from $m$. Then, the integral of the dynamical-friction-force equation can be rewritten using $\d x = \d\beta/R_{\srm}$, which gives
\begin{equation}
    \bm{F}_{\rm DF}
        \propto
            -\,m^{2} \frac{\bm{v}}{|v|} 
            \int_{\tilde{d}_{\crm}}^{\tilde{d}_{\rm max}}\mathrm{d}x\;
            x
            \left[
                1 - \cos\big( \theta[ x ] \big)
            \right]
            .
\end{equation}
Then it is clear that the dynamical-friction dependence on mass is $\Delta E_{\rm DF}\propto m^{2}$. Since this is explicitly true also for all other energy loss mechanisms, it follows that $\Delta E_{\rm tot} \propto m^{2}$. 

By using the approximation $v_{i} \ll v_{\star}$ (which is the regime of interest for our investigation), it is also possible to show that for different initial velocities $v_{1}$ and $v_{2}$ the path traced by the PBH (and therefore the associated energy loss) is approximately the same if the corresponding impact parameters $b_{1}$ and $b_{2}$ satisfy the relation $b_{1}/b_{\crm}( v_{1} ) = b_{2}/b_{\crm}( v_{2} )$. Noticing that for $v_{\star}/v \gg 1$, it follows that $b_{\crm}( v ) \approx R v_{\star}/v$ and $e \approx 1$, the GW energy loss can be approximated as
\begin{equation}
     \Delta E_{\rm GW}
        \approx
            \frac{8}{15}\.
            \frac{m^{2}}{M}\.
            v_{\star}^{7}
            \left(
                \frac{b}{b_{\crm}}
            \right)^{\!-7}
            p(1)
            \, ,
\end{equation}
which is invariant under the transformation described above. Similarly, one can approximate the relevant interior path parameters in the form
\begin{equation}
    \phi_{0}
        \approx
            \pi
            -
            \arccos\!
            \left[
                \left(
                    \frac{b}{b_{\crm}}
                \right)^{\!2}-1
            \right]
            ,
\end{equation}
and
\begin{equation}
    \alpha_{\pm}
        \approx
            R\.\sqrt{\frac{3}{2}\left(1 \pm \sqrt{1-\left( \frac{2b}{3b_{\crm}}\right)^{\!2}} \right)}
            \, .
\end{equation}
Similarly the equation describing the interior path in the case of a realistic NS
\begin{equation}
    \frac{\mathrm{d}r}{\mathrm{d}\phi}
        =
            r^{2}
            \sqrt{
                \frac{v(r)}{R^{2}\.v_{\star}^{2}}
                \left(
                    \frac{ b_{\crm} }{ b }
                \right)^{2}
                -
                \frac{1}{r^{2}}
            }
            \, ,
\end{equation}
is also approximately invariant under this transformation. Since all other parameters describing the path of the neutron star inside are derived from the ones in the equations above. Then the whole internal path is approximately invariant under the transformation $b_{1}/b_{\crm}( v_{1} ) \rightarrow b_{2}/b_{\crm}( v_{2} )$. This was confirmed numerically and can be used to simplify calculations significantly by using the transformation laws described above to generalize one energy loss computation to arbitrary PBH mass and initial velocity respecting the conditions discussed before.
\vfil